\documentclass[a4paper,onecolumn,11pt,unpublished]{quantumarticle}
\pdfoutput=1
\usepackage[utf8]{inputenc}
\usepackage[english]{babel}
\usepackage[T1]{fontenc}
\usepackage{amsmath}
\usepackage{hyperref}
\usepackage{qcircuit}
\usepackage{tikz}
\usepackage{lipsum}
\usepackage{multirow}
\usepackage{graphicx}
\usepackage{subfig}
\usepackage{stmaryrd}
\usepackage{braket}

\DeclareMathOperator{\iSWAP}{\mathsf{iSWAP}}
\DeclareMathOperator{\CZ}{\mathsf{CZ}}
\DeclareMathOperator{\CNOT}{\mathsf{CNOT}}
\DeclareMathOperator{\CXSWAP}{\mathsf{CXSWAP}}
\DeclareMathOperator{\SWAP}{\mathsf{SWAP}}

\begin{document}

\title{Louvre: Relaxing Hardware Requirements of Quantum LDPC Codes by Routing with Expanded Quantum Instruction Set}

\author{Runshi Zhou}
\affiliation{Department of Computer Science and Technology, Tsinghua University, Beijing 100084, China}
\author{Fang Zhang}
\affiliation{Zhongguancun Laboratory, Beijing, China}
\orcid{0000-0002-0000-7101}
\author{Hui-Hai Zhao}
\affiliation{Zhongguancun Laboratory, Beijing, China}
\author{Feng Wu}
\affiliation{Zhongguancun Laboratory, Beijing, China}
\email{wufeng@iqubit.org}
\author{Linghang Kong}
\affiliation{Zhongguancun Laboratory, Beijing, China}
\email{linghang@iqubit.org}
\orcid{0000-0002-5854-5340}
\author{Jianxin Chen}
\affiliation{Department of Computer Science and Technology, Tsinghua University, Beijing 100084, China}
\email{chenjianxin@tsinghua.edu.cn}
\orcid{0000-0002-9365-776X}

\maketitle

\begin{abstract}
  Generalized bicycle codes (GB codes) represent a promising family of quantum low-density parity-check codes, characterized by high code rates and relatively local qubit connectivity. A subclass of the GB code called bivariate bicycle codes (BB codes) has garnered significant interest due to their compatibility with two-layer connectivity architectures on superconducting quantum processors. However, one key limitation of BB codes is their high qubit connectivity degree requirements (degree 6), which exacerbates the noise susceptibility of the system. Building on the recent progress in implementing multiple two-qubit gates on a single chip, this work introduces Louvre---a routing-based framework designed to reduce qubit connectivity requirements in GB codes. Specifically, Louvre-7 achieves degree reduction while preserving the depth of the syndrome extraction circuit, whereas Louvre-8 further minimizes the connectivity by slightly increasing the circuit depth. When applied to BB codes, these two schemes could reduce the average degree to 4.5 and 4, respectively. Crucially, Louvre eliminates some of the long-range, error-prone connections, which is a distinct advantage over prior approaches. Numerical simulations demonstrate that Louvre-7 has an indistinguishable logical error rate as the standard syndrome extraction circuits of GB codes, while Louvre-8 only incurs a slight error rate penalty. Furthermore, by reordering some of the gates in the circuit, we can reduce the coupler length without degrading the performance. Though most of our analysis focuses on GB codes defined on periodic boundary conditions, we further discuss the adaptability of Louvre to open-boundary lattices and defect-containing grids, underscoring its broader applicability in practical quantum error correction architectures.
\end{abstract}

\section{Introduction}

Qubits are susceptible to noise, which limits the width and depth of quantum circuits that can be executed reliably. Thus, quantum error-correction code (QECC) is needed to reduce the error rate by introducing redundancy, encoding a few logical qubits into many physical qubits. The most commonly used type of QECC is known as the stabilizer code \cite{gottesman1997stabilizer}, which is defined as the mutual eigenspace of many Pauli operators called stabilizers. A syndrome extraction circuit (SEC) is designed to measure whether the encoded quantum state still resides in the eigenspace of the stabilizers, and such circuits are frequently run during the quantum computation process to facilitate quantum error correction. In order to run the SEC efficiently, the quantum hardware platform should be equipped with the gates in the circuit as native gates, and connectivity requirements should also be satisfied.

Superconducting quantum hardware \cite{engineer} is one of the most promising quantum computing platforms and has made significant progress in the past several decades \cite{acharya2024quantum}. Its main advantages include fast quantum gates, good scalability, and overlap with traditional semiconductor technology. One of its limitations is the qubit connectivity, as in most of the experiments, superconducting qubits can only be placed on a 2D surface, and only nearest-neighbor connections are allowed. With this constraint, the most well-studied QECC for superconducting qubits is the surface code \cite{fowler2009high,fowler2012surface}. It has a high fault tolerance threshold and only requires 2D grid connectivity with vertex degree 4. A variant of the surface code \cite{McEwen2023relaxinghardware} even further reduced the requirement to hexagonal connectivity with degree 3. The main drawback of surface code is the significant overhead it introduces. It is estimated that \cite{fowler2012surface} hundreds or even thousands of physical qubits are needed to encode a single logical qubit, so that millions of physical qubits might be required for meaningful quantum computation.

The quantum low-density parity-check code (LDPC code) \cite{breuckmann2021quantum} is a broader family of QECC that contains codes with much higher code rates. A general quantum LDPC code often requires complicated connectivity beyond a grid, but a family of quantum LDPC codes called generalized bicycle codes (GB codes) \cite{viszlai_matching_2024} only has finite-length long-range couplers and might be implementable on hardware. Such connectivity requirements could be satisfied with the flip-chip geometry implemented by bump-bonding \cite{rosenberg20173d,field2024modular,kosen2024signal,norris2025performancecharacterizationmultimodulequantum} and the use of superconducting through-silicon vias (TSVs) \cite{yost2020solid,mallek2021fabricationsuperconductingthroughsiliconvias,hazard2023characterization}.
The bump-bonding bonds two chips in a flip-chip geometry, enabling signal routings between two face-to-face surfaces, while TSVs are vertical superconducting interconnects via holes etched through the silicon interposer, enabling signal routings between its top and bottom surfaces. The combination of both techniques enables routing across multiple layers, facilitating the realization of long-range couplers without crossing with other components. 

More interestingly, a subclass of GB codes called bivariate bicycle codes (BB codes) \cite{bravyi_high-threshold_2024} was recently shown to have qubit connections with thickness 2, i.e., the edges in the corresponding graph could be partitioned into two groups and each could be embedded into a 2D plane without crossings between the edges. One group of the edges forms the 2D square grid geometry, while the other contains long-range edges. The drawback of this biplanar implementation of BB codes is that each qubit has degree 6, including 4 short-range couplers and 2 long-range couplers, and such a high degree could lead to fidelity loss. A carefully designed SEC of BB codes \cite{Shaw2025lowering} could reduce the degree to 5 by removing a short-range coupler, but given that a long-range coupler could lead to more infidelity than a short-range coupler, the improvement may not be significant.

In this work, we propose Louvre, which is a family of SECs for GB codes that can remove some of the long-range links and further reduce the degree of the qubits. There are two schemes, Louvre-7 and Louvre-8, named after the depths of the circuits when applied to BB codes. For Louvre-7, which is a depth-7 circuit that has the same depth as the original BB code SEC, the qubit degree is 4.5 on average, with 0.5 long-range connection and 1 short-range connection removed. By slightly increasing the depth to 8, the Louvre-8 circuit can further reduce the degree of qubit to 4, removing 1 long-range connection and 1 short-range connection on average. We have also shown that Louvre has a level of logical error rate comparable to that of the original SEC of BB codes. We conducted numerical studies and found that Louvre-7 has almost the same logical error rate as the original BB code SEC, while Louvre-8 only slightly increases the error rate. 

Furthermore, reordering of some of the gates in the circuit enables reduction in the length of the long-range couplers. We introduce Louvre-7R and Louvre-8R, which are optimized variants of Louvre-7 and Louvre-8 that can potentially achieve this reduction without compromising the logical error rate performance. While the coupler length itself might not dictate the complexity of hardware implementation, as the couplers might not be arranged in a straight line, our numerical analysis demonstrates that shortened couplers could mitigate hardware requirements under some realistic metrics.

The main idea behind Louvre is to route the ancilla qubits without introducing additional overhead. Note that a $\CNOT$ gate followed by a $\SWAP$ gate is $\CXSWAP$, which is equivalent to an $\iSWAP$ gate up to single-qubit gates. In the rest of the work, we will use $\iSWAP$ and $\CXSWAP$ interchangeably because single-qubit gates have much lower error rate and shorter gate time compared to two-qubit gates. A conventional SEC consists of $\CNOT$ gates between ancilla qubits and the data qubits associated with them, so by changing the $\CNOT$ gates to $\iSWAP$ gates, we can have free $\SWAP$ gates to move the qubits. Louvre takes advantage of this feature and reuses some of the connections in the circuits, so that some other connections could be removed. Louvre-7 requires $\iSWAP$ and $\CNOT$ as the two-qubit native gates, while Louvre-8 requires $\iSWAP$, $\CNOT$ and $\SWAP$. Louvre-7R and Louvre-8R could replace the $\CNOT$ gates by $\iSWAP$, so that Louvre-7R requires $\iSWAP$ only, and Louvre-8R requires $\iSWAP$ and $\SWAP$. These gates could be achieved with high fidelity on a superconducting device, given the AshN scheme proposed in \cite{chen_one_2024} and experimentally tested in \cite{chen2025efficientimplementationarbitrarytwoqubit}.

The rest of this work is structured as follows. In Section~\ref{sec:background}, we introduce the background of quantum error correction codes and the extended gate set of superconducting qubits used in our construction. In Section~\ref{reduce_number}, we propose the schemes Louvre-7 and Louvre-8, which can reduce the number of couplers in the device. In Section~\ref{reduce_length}, we present Louvre-7R and Louvre-8R, which further reduce hardware requirements by decreasing the coupler length. The numerical results on the logical error rates of our construction, together with the effects of shortened couplers, can be found in Section~\ref{sec:numerics}. Finally, we conclude this work with discussions in Section~\ref{sec:discussion}.

\emph{Note added:}
The idea of cost-free routing of qubits with $\iSWAP$ gates was first proposed in another work by the authors \cite{zhou2024halmaroutingbasedtechniquedefect}, in the context of implementing the surface code on a 2D grid with defects. A recent independent work \cite{geher2025directionalcodesnewfamily} constructed a new family of LDPC codes called directional codes on a 2D grid based on similar ideas.

\section{Background}
\label{sec:background}
\subsection{Generalized Bicycle code}


The generalized bicycle code~\cite{viszlai_matching_2024} is a class of stabilizer codes constructed by a pair of generating polynomials, $A$ and $B$, which define the shape of the stabilizers, that is, the relative positions of the data qubits from which an ancilla qubit need collect error syndromes. Each of the generating polynomials $A$ and $B$ consists of multiple terms $A_i$ and $B_i$, and each term corresponds to the relative position of a data qubit in the stabilizer with respect to the associated ancilla qubit. When we fix both the number of terms in $A$ and $B$ to 3, we get a bivariate bicycle code~\cite{bravyi_high-threshold_2024}, which is a subclass of GB codes that is considered more suitable for superconducting hardware. 

\begin{figure}[h]
\centering
\includegraphics[width=  0.6
\columnwidth]{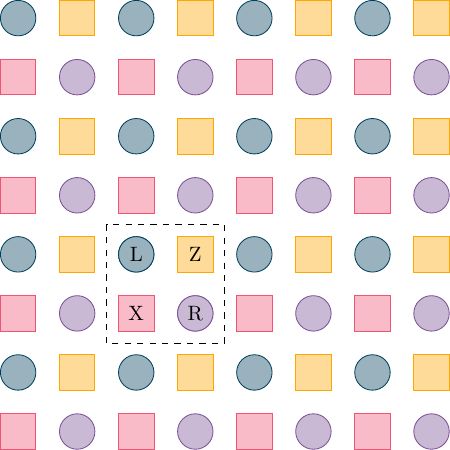} 
\caption{The bulk region of a generalized bicycle code with a basic unit in the boxed area.}
\label{Basic_Unit}
\end{figure}

A typical set-up of a GB code is shown in Fig.~\ref{Basic_Unit}, with periodic boundary conditions assumed, in another word, on the topology of a torus. The code is made up of $l$ rows time $m$ columns of \textit{basic units}, indicated as the boxed area. A basic unit comprises 4 physical qubits, which are usually labeled as \textbf{L, R, X, Z} standing for left and right data qubits, and X and Z-ancilla qubits respectively, as shown in Fig.~\ref{Basic_Unit}. It is helpful to see the total qubit configuration as the composition of four sublattices, each containing all qubits of a certain type. Thus, there would be a left (right) data qubit sublattice, and an X-ancilla (Z-ancilla) sublattice. Then, a basic unit contains one qubit from each sublattice.

\begin{figure}[h]
\centering
\subfloat[Toric code]{\includegraphics[width=  0.45\columnwidth]{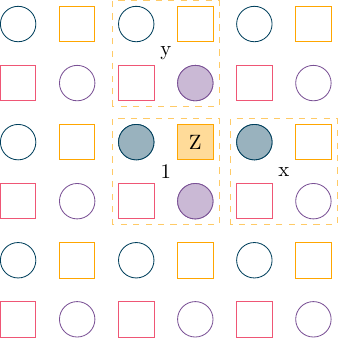}} \hspace{10mm} 
\subfloat[$\llbracket 18,4,4\rrbracket$ BB code]{\includegraphics[width=  0.45\columnwidth]{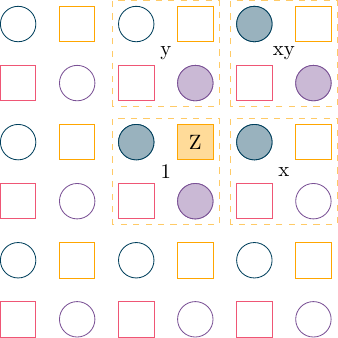}}
\caption{(a) The shape of a Z-stabilizer for the toric code in the language of generalized bicycle code.  Filled circles are data qubits included in the Z-stabilizer, associated with the Z-ancilla denoted by the filled square (B) A $\llbracket18,4,4\rrbracket$ BB code as example.}
\label{surf_demo}
\end{figure}

As an example, to see which data qubits an Z-ancilla needs to interact with during syndrome extraction, we first look at the generating polynomial $A$. For a Z-ancilla qubit, each term $A_i$ in $A$ corresponds to a right data qubit. $A_i=x^\alpha y^\beta$ means a Z-ancilla needs to interact with the right data qubit in the basic unit that is, from the basic unit in which the Z-ancilla is at, $\alpha$ blocks to its right and $\beta$ blocks up. Likewise, terms in $B$ correspond to left data qubits in the given basic units, following the same rule of positioning. As an example, in this language, Kitaev's toric code has generating polynomials $A = 1 + y$,  $B = 1 + x$, as shown in Fig.~\ref{surf_demo}(a).

For an X-ancilla, however, $A_i$ and $B_i$ correspond to left and right data qubits, respectively, and $A_i=x^\alpha y^\beta$ is the basic unit that is $\alpha$ blocks to its left and $\beta$ blocks down. So, it is easy to see that, for the GB code on the periodic boundaries, all the X(Z)-stabilizers that are measured during syndrome extraction share the same shape. That is, the relative positions between the data qubits in each measured X(Z)-stabilizer and the associated ancilla qubit are the same. The shape of a X-stabilizer is the same as the shape of a Z-stabilizer by rotation of $180^\circ$.

For clarity in the following discussion, we define the following terminologies.
\begin{enumerate}
    \item $q_{A_i,X_j}$ indicates the data qubit corresponding to the term $A_i$ for the X-ancilla qubit $X_j$. More colloquially, we will refer to it as an X-ancilla and its $A_i$.
    \item $v_{A_i,X}$ indicates the vector that points from any X-ancilla $X_j$ to the data qubit $q_{A_i,X_j}$.
\end{enumerate}
As we have previously discussed, $q_{A_i,X_j}$ and $q_{B_i,Z_j}$ are left data qubits, while $q_{B_i,X_j}$ and $q_{A_i,Z_j}$ are right data qubits.

\subsection{Syndrome Extraction Circuit}
When performing syndrome extraction for the generalized bicycle code, we usually assume that all operations are global: At each layer, the actions of all X-ancilla qubits are the same ($\CNOT$ with the data qubit at a particular relative position, for example), and so are those of the Z-ancilla qubits. 

To ensure commutation, a global syndrome extraction circuit typically consists of three phases.
\begin{itemize}
    \item In Phase 1, all X-ancilla qubit interacts with their $A_i\in F_x$ and all Z-ancilla qubit interacts with their $A_i\in F_z$ in arbitrary order, with $F_x \cap F_z = \emptyset$ and $F_x \cup F_z$ including all terms in $A$.
    \item In Phase 2, all ancilla qubits interacts with all their $B_i$ in arbitrary order.
    \item In Phase 3, all X-ancilla qubit interacts with their $A_i\in F_z$ and all Z-ancilla qubit interacts with their $A_i\in F_x$ in arbitrary order.
\end{itemize}

For example, a global instruction of a syndrome extraction circuit for a BB code is shown in Table~\ref{regular_BB}, with $A_1,A_2\in F_x$ and $A_3\in F_z$.

\begin{table}[h]
\small
\begin{center}
\caption{An example instruction for a global syndrome extraction circuit of a BB code. $A_1$ means at this layer, all X-ancilla $X_i$ (or Z-ancilla) interact with $q_{A_1,X_i}$.}
\label{regular_BB}
\begin{tabular}{ |c|c|c|c|c|c|c|c| } 
 \hline
Phase & \multicolumn{2}{|c|}{1} & \multicolumn{3}{|c|}{2} & \multicolumn{2}{|c|}{3} \\
 \hline
Layer & 1&2&3&4&5&6&7 \\ 
\hline
X-ancilla & $A_1$ & $A_2$ & $B_1$ & $B_2$ & $B_3$ & $A_3$ &  \\ 
\hline
Z-ancilla &  & $A_3$ & $B_1$ & $B_2$ & $B_3$ & $A_2$ & $A_1$  \\ 
 \hline
\end{tabular}
\end{center}

\end{table}

To see why this syndrome extraction circuit works, let us assume that an X-ancilla $X_0$ and a Z-ancilla $Z_0$ have some non-trivial overlap in the data qubits in their associated stabilizers.

Say there is an overlapping left data qubit $q_{A_i,X_0}$, the data qubit $A_i$ of $X_0$, which happens to be the same qubit as $q_{B_j,Z_0}$, $B_j$ of $X_0$. Let us call this qubit $d_1$. This means $v_{A_i,X}-v_{B_j,Z}$ points from $X_0$ to $Z_0$. Then $v_{A_i,X}-v_{B_j,Z} = -v_{A_i,Z}+v_{B_j,X}$ also points from $X_0$ to $Z_0$. Thus, there must be an overlapping right data qubit $q_{B_j,X_0}$, which is the same as $q_{A_i,Z_0}$. In other words, $B_j$ of $X_0$ must also be $A_i$ of $Z_0$. Let us call this qubit $d_2$. Then, it is apparent to see that overlapping data qubits always come in pairs.

If $A_i \in F_x$, $X_0$ interacts with $d_1$ in Phase 1, while $Z_0$ interacts with $d_1$ in Phase 2. In addition, $X_0$ interacts with $d_2$ in Phase 2, whereas $Z_0$ interacts with $d_2$ in Phase 3. So both $d_1$ and $d_2$ interact with $X_0$ before they interact with $Z_0$. Similarly, if $A_i \in F_z$, both $d_1$ and $d_2$ interact with $Z_0$ before they interact with $X_0$.

Hence, there is always an even number of data qubits that interact with $X_0$ before $Z_0$, and an even number of data qubits that interact with $Z_0$ before $X_0$. This guarantees that commutation is preserved \cite{geher2023tangling}.

\subsection{Expanded Native Instruction Set}

In this work, we proceed from the premise that the quantum chip in use can natively implement multiple types of two-qubit gates, including $\CNOT$, $\SWAP$, and $\iSWAP$ gates (the latter being equivalent to a $\CNOT$ gate followed by a $\SWAP$ gate). This assumption is valid, particularly for mainstream tunable coupler transmon architectures supporting both XX and YY interactions, where the AshN gate scheme proposed in~\cite{chen_one_2024} enables the high-fidelity realization of arbitrary two-qubit gates. Due to differences in gate duration, we adopt the hardware parameters described in~\cite{chen2025efficientimplementationarbitrarytwoqubit}, under which $\SWAP$ gates exhibit a $1.5$-fold higher gate error than $\CZ$ and $\iSWAP$ gates when decoherence is the dominant factor. We will investigate how varying this factor impacts Louvre's performance in subsequent simulations.

\section{Reducing the Number of Couplers}
\label{reduce_number}
\begin{figure}[h]
\centering
\includegraphics[width=  0.6
\columnwidth]{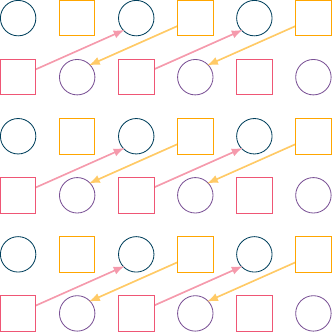} 
\caption{$v_{B_i,Z}$ (red arrow) and $v_{B_i,X}$ (blue arrow).}
\label{same_vector}
\end{figure}

In this section, we begin introducing the method of Louvre. The primary goal of Louvre is to reduce the number of couplers required to perform the syndrome extraction for GB code, thus relaxing the connectivity requirement on quantum hardware. The key insight behind Louvre is that the vector $v_{A_i,Z} = -v_{A_i,X}$. In other words, while $v_{A_i,X}$ points from an X-ancilla qubit to its $A_i$, it also points from the $A_i$ of a Z-ancilla qubit to that Z-ancilla, as shown in Fig.~\ref{same_vector}. Since couplers do not have an inherent direction, we can potentially let the same coupler facilitate both of the interactions, with the help of some qubit routing. 

\begin{figure}[h!]
\centering
\subfloat[]{\includegraphics[width=  0.3\columnwidth]{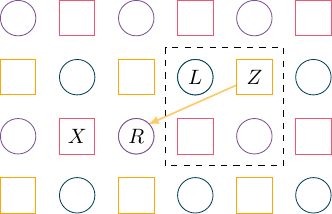}} \hspace{6mm} 
\subfloat[]{\includegraphics[width=  0.3\columnwidth]{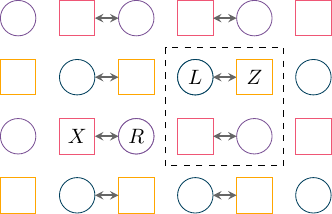}} \hspace{6mm} 
\subfloat[]{\includegraphics[width=  0.3\columnwidth]{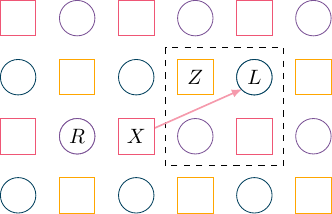}}\\
\subfloat[]{
\hspace{2mm}
\Qcircuit@C=1em@R=1.5em{
\lstick{X}&\qw      &\qswap     &\qw      &\qw&R \\
\lstick{R}&\ctrl{1}&\qswap\qwx  &\ctrl{1}&\qw&X \\
\lstick{Z}&\targ    &\qswap     &\targ    &\qw&L \\
\lstick{L}&\qw      &\qswap\qwx &\qw      &\qw&Z \\
 }
}
\caption{(a-c) The 2D diagram and (d) the quantum circuit of an example reusing the same coupler by inserting a $\SWAP$ gate in between.}
\label{Ldemo}
\end{figure}

For example, after performing two-qubit gates between all Z-ancilla and their $A_i$, we can swap the location of the Z-ancilla with left data qubits and the X-ancilla with right data qubits (typically the adjacent ones), via $\CXSWAP$ or $\SWAP$ gates. Afterward, the same couplers can be used to perform two-qubit gates between all X-ancilla and their $A_i$, as shown in Fig.~\ref{Ldemo}. The action of swapping all qubits in adjacent rows or columns resembles flipping louvre blinds, so is this method named after. Depending on our goals, there are several ways we could apply this technique, leading to multiple types of circuits each with different features, as demonstrated in the following sections.

\subsection{Louvre-7}
Louvre-7 is the most basic circuit, which reduces the number of couplers without any overhead in noise or circuit depth. We will walk through the procedure for building a Louvre-7 circuit, with the $\llbracket18,4,4\rrbracket$ BB code as an example~\cite{wang_demonstration_2025}, whose generating polynomial is $A = 1 + y +xy$, $B= 1 +x + xy$. The gate instruction for this BB code is shown in Table~\ref{L7order}, with the shape of the stabilizers shown in Fig.~\ref{surf_demo}(b).

\begin{table}[h]
\small
\begin{center}
\caption{The Louvre-7 syndrome extraction circuit gate instruction for the $\llbracket18,4,4\rrbracket$ BB code. In the table, $A_i$ means all ancilla of that type perform $\CNOT$ gates with their corresponding $A_i$, unless the gate is otherwise specified in the following parenthesis.}
\label{L7order}
\begin{tabular}{ |c|c|c|c|c|c|c|c| } 
 \hline
 Phase & \multicolumn{2}{|c|}{1} & \multicolumn{3}{|c|}{2} & \multicolumn{2}{|c|}{3} \\
 \hline
Layer & 1&2&3&4&5&6&7 \\ 
\hline
X-ancilla & $A_1$ & $A_2$ & $B_2$ & $B_3$ & $B_1$($\CXSWAP$) & $A_3$ &  \\ 
\hline
Z-ancilla &  & $A_3$ & $B_2$ & $B_3$ & $B_1$($\CXSWAP$) & $A_2$ & $A_1$  \\ 
 \hline
\end{tabular}

\end{center}
\end{table}

First, we examine the generating polynomials $A$ and $B$ to see which polynomial contains more terms. Without loss of generality, we will assume that $A$ contains more terms than $B$ (if not, we flip the code by the diagonal and thus swap $A$ and $B$). Then, we choose a term in $B$. Again, without loss of generality, we assume this term is $B_1$.

\begin{figure}[h!]
\centering
\subfloat[Phase 1]{\includegraphics[width=  0.3\columnwidth]{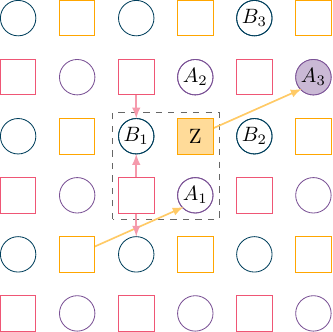}} \hspace{5mm} 
\subfloat[Phase 2 layer 3-4]{\includegraphics[width=  0.3\columnwidth]{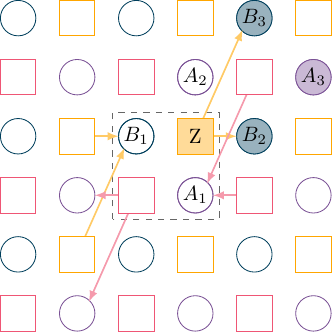}} \hspace{5mm} 
\subfloat[Phase 2 layer 5]{\includegraphics[width=  0.3\columnwidth]{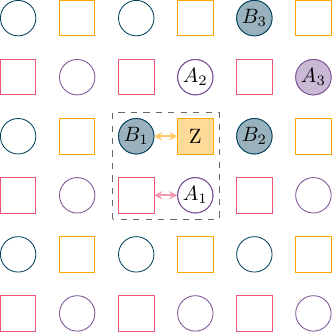}}\\
\subfloat[Phase 3]{\includegraphics[width=  0.3\columnwidth]{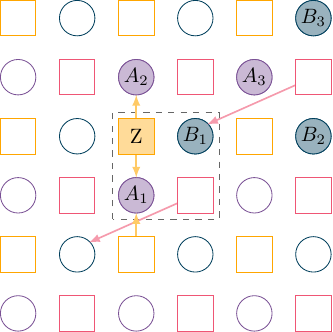}}\hspace{5mm} 
\subfloat[Connectivity]{\includegraphics[width=  0.3\columnwidth]{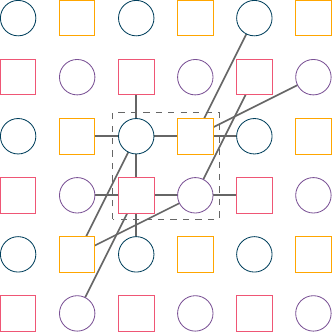}}
\caption{The Louvre-7 syndrome extraction circuit of the $\llbracket18,4,4\rrbracket$ BB code focusing on one particular basic unit.}
\label{L7Figs}
\end{figure}

In our BB code example, both $A$ and $B$ have 3 terms, so there is no need to adjust, and $B_1 = 1$. Phase 1 of Louvre-7 is the same as a regular syndrome extraction circuit, as shown in Fig.~\ref{L7Figs}(a), where we focus on the boxed basic unit and use arrows to denote the two-qubit gates that occur.

During Phase 2 of the syndrome extraction circuit, we want to perform some qubit routing, so that the gates in Phase 3 can use the same couplers as the gates in Phase 1. Therefore, we perform a $\CXSWAP$ gate, instead of the regular $\CNOT$ gate, between all ancilla qubits and their $B_1$. The $\CXSWAP$ gate facilitates qubit routing without introducing additional noise or circuit depth. The circuit for Phase 2 is shown in Fig.~\ref{L7Figs}(b-c).

The $\CXSWAP$ gates change the qubit configurations. Under periodic boundary condition, all qubits of the same type perform the same gate in each layer. Therefore, after the $\CXSWAP$ layer between all X-ancilla and their $B_1$, all X-ancilla move by $v_{B_1,X}$, while all the right data qubits move by $-v_{B_1,X}$. In the language of sublattices, the X sublattice moves by $v_{B_1,X}$ which is 1 grid distance to the right in our example, while the right data qubit sublattice moves by $-v_{B_1,X}$, one grid distance to the left. 

Consequently, the relative positions between the X-ancilla and right data qubits are different before and after the routing takes place, and the same is true for the routing layer between the Z-ancilla and left data qubits. In our example, the right data qubit that used to be on the left of an X-ancilla is at 3 grid distances away after routing. By leveraging the degree of freedom of ordering the two-qubit gates in the gate instruction, the change in qubit configuration could potentially shorten some long-range interaction. But for clarity in the current discussion, we will perform the routing in the last layer of Phase 2.

Additionally, by the end of Phase 2, when all routing is completed, we see that the routing layer between the Z-ancilla and left data qubits also moves the two sublattices in a similar way: the Z-ancilla sublattice moves by $v_{B_1,Z} = -v_{B_1,X}$, while the left data qubit sublattice move by $-v_{B_1,Z} = v_{B_1,X}$. As a result, the relative positions between the X-ancilla sublattice and the left data qubit sublattice are the same as before in Phase 1, since they both move by $v_{B_1,X}$ during Phase 2. Therefore, the relative position between an X-ancilla qubit and any of its $A_j$ is still $v_{A_j,X}=-v_{A_j,Z}$. The same is true between the Z-ancilla sublattice and the right data qubit sublattice, which both move by $-v_{B_1,X}$ during the routing layer. 

\begin{figure}[h]
\centering
\includegraphics[width=  0.4
\columnwidth]{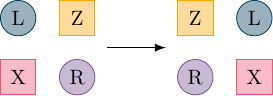} 
\caption{The qubit configuration in a basic unit before and after routing.}
\label{L7BU}
\end{figure}

Meanwhile, the qubit configuration in each basic unit changed as shown in Fig.~\ref{L7BU}. Since the X-ancilla and the left data qubits have switched positions with the right data qubits and the Z-ancilla respectively, we can reuse the same couplers in Phase 1 to facilitate the gates in Phase 3, as shown in Fig.~\ref{L7Figs}(d). 

After a regular round of syndrome extraction circuit, we follow it by a \textit{reversed round}, where we perform all the gates in the regular circuit in reversed order, except the reset and measurement operations. This round will effectively extract error syndrome while returning all the qubits to their original configuration, so that we can cycle between regular rounds and reversed rounds thereafter. 

Compared to the original degree of connection, which is the total number of terms in $A$ and $B$, using the Louvre-7 circuit, the average degree requirement is reduced to
\begin{equation}
    d_{L7} = n_a + n_b - \frac{1}{2}\max(n_a,n_b)
\end{equation}
where $n_a$ and $n_b$ are number of terms is $A$ and $B$ respectively. This is particularly helpful when there are significantly more terms in $A$ and only a small number of terms in $B$, or vice versa. An example would be the $\llbracket72,8,9\rrbracket$ GB code~\cite{lin_quantum_2023}, which has 2 terms in $A$ and 6 terms in $B$.

For the BB code in our example, as well as other BB codes, the resulting average degree is 4.5, as shown in Fig.~\ref{L7Figs}(e). The connections required by each ancilla in the configuration before the syndrome extraction circuit is listed in Table~\ref{L7Con}.

\begin{table}[h]
\small
\begin{center}
\caption{The required connection for each ancilla at layer 1 in the Louvre-7 circuit. The terms from generating polynomials stand for the corresponding data qubits of that ancilla.}
\label{L7Con}
\begin{tabular}{ |c|c| } 
\hline
X-ancilla & $A_1,A_2,B_1,B_2,B_3$ \\ 
\hline
Z-ancilla & $A_3,B_1,B_2,B_3$ \\ 
\hline
\end{tabular}
\end{center}
\end{table}

\subsection{Louvre-8}

Louvre-8 aims to further reduce the number of couplers while tolerating some overhead in noise and circuit depth. We will walk through its procedure with the same $\llbracket18,4,4\rrbracket$ BB code as an example. The gate instruction of its syndrome extraction circuit is shown in Table~\ref{L8order}.

\begin{table}[h]
\small
\begin{center}
\caption{The Louvre-8 syndrome extraction circuit gate instruction for the $\llbracket18,4,4\rrbracket$ BB code.}
\label{L8order}
\begin{tabular}{ |c|c|c|c|c|c|c|c|c| } 
 \hline
  Phase & \multicolumn{2}{|c|}{1} & \multicolumn{4}{|c|}{2} & \multicolumn{2}{|c|}{3} \\
 \hline
Layer & 1&2&3&4&5&6&7&8 \\ 
\hline
X-ancilla & $A_2$ & $A_1$ & $B_2$ & $A_1$($\SWAP$)& $B_3$ & $B_1$($\CXSWAP$) & $A_1$($\SWAP$) & $A_3$ \\ 
\hline
Z-ancilla &  & $A_3$ & $B_3$ & $A_1$($\SWAP$)& $B_2$ & $B_1$($\CXSWAP$) & $A_1$($\CXSWAP$) & $A_2$  \\ 
 \hline
\end{tabular}
\end{center}
\end{table}

First, we choose a term in $A$ and a term in $B$, and call them $A_1$ and $B_1$. Typically, we want the gates with $A_1$ and $B_1$ to be short-ranged, to reduce the overall interaction distance. We divide the rest of the terms in $B$ into two sets, $G_x$ and $G_z$. In our example, we have $B_2 \in G_x$ and $B_3 \in G_z$.

\begin{figure}[h!]
\centering
\subfloat[Phase 1]{\includegraphics[width=  0.3\columnwidth]{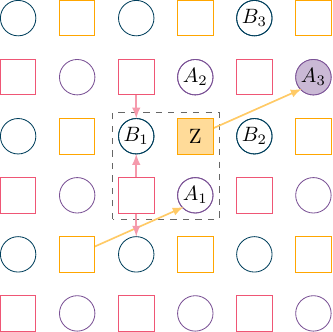}} \hspace{5mm} 
\subfloat[Phase 2 layer 3]{\includegraphics[width=  0.3\columnwidth]{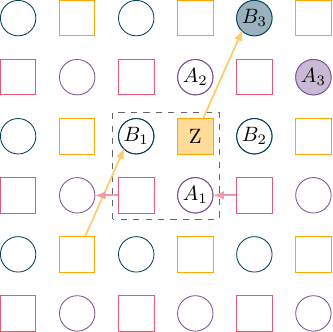}}\hspace{5mm} 
\subfloat[Phase 2 layer 4]{\includegraphics[width=  0.3\columnwidth]{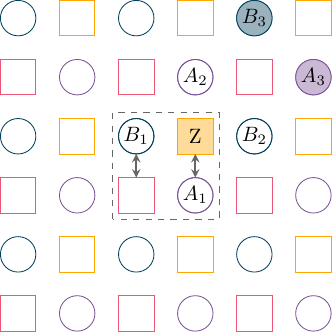}} \\
\subfloat[Phase 2 layer 5]{\includegraphics[width=  0.3\columnwidth]{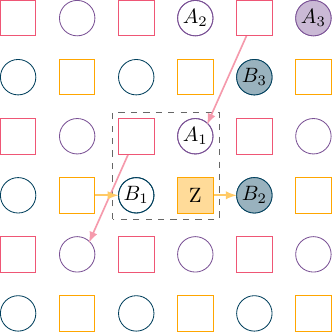}} \hspace{5mm} 
\subfloat[Phase 2 layer 6]{\includegraphics[width=  0.3\columnwidth]{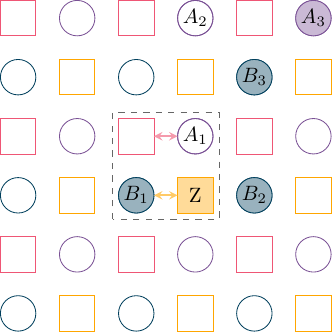}}\hspace{5mm} 
\subfloat[Phase 3 layer 7]{\includegraphics[width=  0.3\columnwidth]{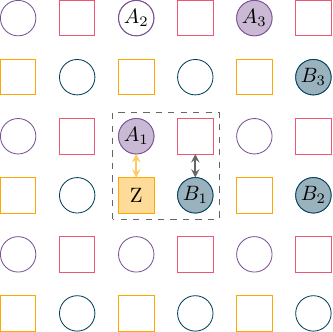}} \\
\subfloat[Phase 3 layer 8]{\includegraphics[width=  0.3\columnwidth]{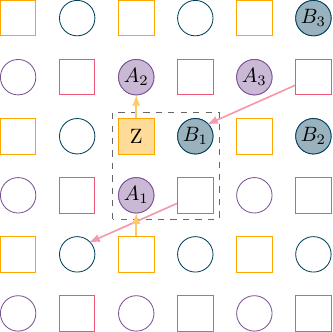}}\hspace{5mm} 
\subfloat[Connectivity]{\includegraphics[width=  0.3\columnwidth]{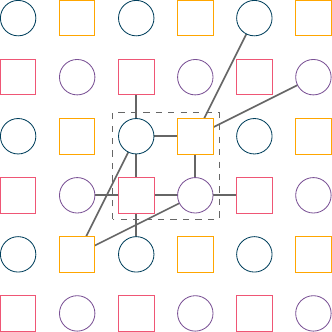}}\hspace{5mm} 
\caption{The Louvre-8 syndrome extraction circuit of the $\llbracket18,4,4\rrbracket$ BB code focusing on one particular basic unit. }
\label{L8Figs}
\end{figure}

We perform Phase 1 the same as before and then divide Phase 2 of the syndrome extraction circuit into two sub-phases, Phase 2A and Phase 2B. In Phase 2A, all X-ancilla interact with their $B_i \in G_x$, while all the Z-ancilla interact with their $B_i \in G_z$, as shown in Fig.~\ref{L8Figs}(b).

Then, we manually insert a routing layer using $\SWAP$ gates, via couplers connecting all ancilla with their corresponding $A_1$, as shown in Fig.~\ref{L8Figs}(c). During this routing layer, the X-ancilla sublattice moves together with the right data qubit sublattice, while the Z-ancilla sublattice moves together with the left data qubit sublattice. Thus, all ancilla qubits preserve their relative positions with all of their $B_j$. This allows us to use the same set of couplers in Phase 2A and 2B.

In Phase 2B, we perform two-qubit gates between each X-ancilla with their $B_i \in G_z$ and between each Z-ancilla with their $B_i \in G_x$, using the same couplers as in Phase 2A. In the final layer of Phase B, we perform $\CXSWAP$ gates between all ancilla qubits with their $B_1$, as shown in Fig.~\ref{L8Figs}(d). During the second routing layer, the X-ancilla sublattice moves together with the left data qubit sublattice, while the Z-ancilla sublattice moves together with the right data qubit sublattice.

It should be clear that, at this point, the relative positions between all four sublattices are shuffled by the two routing layers. So, before Phase 3, we need to reverse the $\SWAP$ routing that was performed between Phase 2A and Phase 2B using another routing layer, so that the relative positions of all ancilla qubtis and all their $A_i$ are restored as in Phase 1. To do that, we need to perform $\SWAP$ gates between all ancilla qubits and their $A_1$ in the first layer of Phase 3. Half of these $\SWAP$ gates can be combined with $\CNOT$ gates and converted to $\CXSWAP$ gates (or vice versa), depending on whether $A_1$ is in $F_x$ or $F_z$. Then, using the same couplers as in Phase 1, we perform two-qubit gates between Z-ancilla with their $A_i \in F_x$ and X-ancilla with their $A_i \in F_z$, excluding $A_1$, which is already interacted with during the first layer of Phase 3.

The interaction between ancilla qubits and all terms in $A$ and in $B$, excluding $A_1$ and $B_1$, which are used as routing layers, is facilitated by shared couplers in the Louvre-8 circuit. Thus, the Louvre-8 circuit will reduce the average degree of the code to 
\begin{equation}
    d_{L8} = \frac{1}{2}(n_a + n_b) + 1,
\end{equation}
where $n_a$ and $n_b$ are the number of terms is $A$ and $B$ respectively. For our example, the reduced connectivity has a degree of 4, and is shown in Fig.~\ref{L8Figs}(h). The connections required by each ancilla in the configuration before the syndrome extraction circuit is listed in Table~\ref{L8Con}. Like in the Louvre-7 circuit, we need to cycle between regular rounds and reversed rounds of the syndrome extraction circuit.

\begin{table}[h]
\small
\begin{center}
\caption{The required connection for each ancilla at layer 1 in the Louvre-8 circuit. The terms from generating polynomials stand for the corresponding data qubits of that ancilla.}
\label{L8Con}
\begin{tabular}{ |c|c| } 
\hline
X-ancilla & $A_1,A_2,B_1,B_2$ \\ 
\hline
Z-ancilla & $A_1,A_3,B_1,B_3$ \\ 
\hline
\end{tabular}
\end{center}
\end{table}

\section{Reducing the Distance of Interaction}
\label{reduce_length}
As we have previously mentioned, by leveraging the degree of freedom of reordering the two-qubit gates within each phase of the syndrome extraction circuit and employing additional layers of $\CXSWAP$ routing, we could reduce the distance of long-range interactions. We view this as a secondary goal of Louvre, to reduce the coupler length, thus further relaxing the requirement imposed on experimental hardware. We introduce Louvre-7R and Louvre-8R, which are optimized variants of Louvre-7 and Louvre-8 with shortened couplers.

Some examples of reduced coupler length could be found in Table~\ref{code_results}, where the definition and source of the codes could be found in Table~\ref{code_parameter}. We evaluate the syndrome extraction circuit of these codes in terms of their averaged degree - the number of couplers a qubit connects to - and averaged total interaction distances - the sum of the distances of all interaction a qubit connects to. The results are presented in Table~\ref{code_results}. When calculating the interaction distance, we use the L1 distance (also known as the Manhattan distance). For example, the distance between $(0,0)$ and $(6,3)$ is $6+3=9$. Note that the circuits here are designed by human ansatz, instead of algorithmic compilation, so some of the results might be suboptimal. The results generalize to other codes with the same stabilizer shape, such as the $\llbracket72,12,6\rrbracket$ and $\llbracket108,8,10\rrbracket$ BB codes.

\begin{table}[h]
\small
\begin{center}
\caption{The averaged degree (left) and the averaged total interaction distances (right) of different codes with different circuits.}
\label{code_results}
\begin{tabular}{ |c|c|c|c|c|c|c|c|c|c|c| } 
 \hline
  $\llbracket n,k,d\rrbracket$& \multicolumn{2}{|c|}{Regular} & \multicolumn{2}{|c|}{Louvre-7}& \multicolumn{2}{|c|}{Louvre-7R}& \multicolumn{2}{|c|}{Louvre-8}& \multicolumn{2}{|c|}{Louvre-8R} \\ 
\hline
\hline
$\llbracket18,4,4\rrbracket$& 6&10&4.5&7.5&-&-&4&6&-&- \\ 
\hline
$\llbracket72,8,9\rrbracket$& 8&54&5&28&5&26&4.5&28&4.5&26 \\ 
\hline
$\llbracket72,8,10\rrbracket$& 8&-&6&-&-&-&5&-&-&- \\ 
\hline
$\llbracket72,12,6\rrbracket$&6&22&4.5&16.5&4.5&13.5&4&12&4.5&10.5 \\ 
\hline
$\llbracket72,8,4\rrbracket$&6&10&4.5&7.5&3.5&3.5&4&6&-&- \\ 
\hline
$\llbracket96,10,12\rrbracket$& 8&62&6&43&6&37&5&32&5.5&27.5 \\ 
\hline
$\llbracket128,16,8\rrbracket$& 8&44&6&32&6&26&5&23&5.5&19.5 \\ 
\hline

\end{tabular}
\end{center}
\end{table}

Although the coupler length itself does not directly correlate with the complexity of hardware implementation (because the coupler may not be arranged in straight lines), as we will see in Section~\ref{subsec:length_numerics}, the reduction of coupler length indeed mitigates hardware requirements under some practical metrics.

\subsection{Ordering}
\label{ordering}
Assume that, during a phase of the syndrome extraction circuit, each Z-ancilla needs to interact with a set of right data qubits, namely their $A_1, A_2$ and $A_3$, but the order in which the interactions take place is arbitrary. Conventionally, we carry out these interactions via $\CNOT$ gates, and the Z-ancilla remains static during this phase of the circuit.

However, if we were to replace the gate between $Z_j$ and a certain $A_i$ with a $\CXSWAP$ gate, then after the $\CXSWAP$ gate, the Z-ancilla sublattice moves by $v_{A_i,Z}$ and the right data qubit sublattice move by $-v_{A_i,Z}$. Thus, the vector that points from a Z-ancilla to its $A_k$ becomes $v_{A_k,Z}'=v_{A_k,Z}-2v_{A_i,Z}$, which could be shorter than $v_{A_k,Z}$. 

\begin{figure}[h!]
\centering
\subfloat[Before Routing]{\includegraphics[width=  0.6\columnwidth]{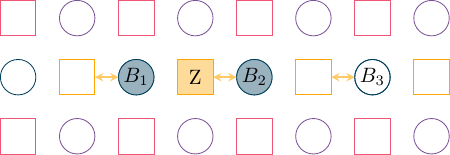}} \\ 
\subfloat[After Routing]{\includegraphics[width=  0.6\columnwidth]{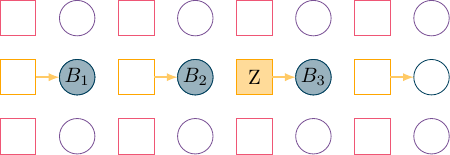}}
\caption{An example of reducing interaction distances via $\CXSWAP$ routing. }
\label{LCP3}
\end{figure}

For example, let us have $v_{A_1,Z} = (-1,0)$, $v_{A_2,Z} = (1,0)$ and $v_{A_3,Z} = (3,0)$ in units of the grid distance. Conventionally, to implement the two-qubit gate between $Z_j$ and its $A_3$, a coupler will be needed to directly connect the two qubits, and the coupler will at least have a length of 3. However, if we replace the $\CNOT$ gate between Z-ancilla and their $A_2$ with a $\CXSWAP$ gate, then the Z-ancilla sublattice will move to the right by 1, while the right data qubit sublattice will move to the left by 1, as shown in Fig.~\ref{LCP3}. Then, the relative position between the X-ancilla and their $A_3$ becomes $(3,0) - 2\times(1,0) = (1,0)$, directly adjacent to the Z-ancilla. Thus the distance of interaction is shortened. Meanwhile, after the $\CXSWAP$ gate, $v_{A_1,Z}'$ becomes $(0,-3)$, which is longer than the original $v_{A_1,Z}$. Therefore, we want the Z-ancilla to interact with their $A_1$ before the $\CXSWAP$ gate is performed with its $A_2$, and then have them interact with their $A_3$.

The best way of applying this technique is usually apparent for a code with small weight, such as BB code, whereas for large weight GB code, it might require an algorithmic search for the optimal ordering and initial configuration. Regardless, we expect to obtain a gate instruction in the general form as follows: 
\begin{equation}
    F_1 \to A_a(\CXSWAP)\to F_2\to A_b(\CXSWAP)\to F_3\to A_c(\CXSWAP)...
\end{equation}
where $F_i$ are sets of terms in $A$, standing for non-routing sections where $\CNOT$ gates are performed between ancilla qubits and all of their corresponding $A_j\in F_i$ in arbitrary order. In between non-routing sections $F_1, F_2, \ldots$ are routing layers $A_a, A_b, \ldots$, where ancilla qubits performs $\CXSWAP$ gates with its data qubit of the specified term.

\subsection{Louvre-7R}
\begin{figure}[h]
\centering
\includegraphics[width=  0.6
\columnwidth]{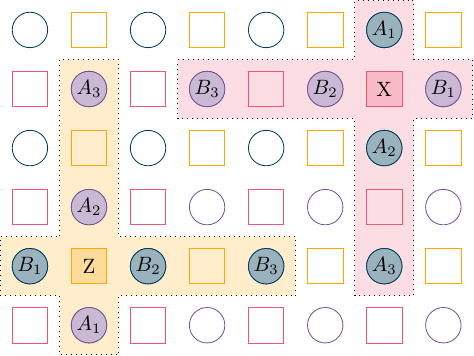} 
\caption{The shape of the stabilizers in the $\kappa=2$ La-Cross code, indicated by the filled data qubit in the shaded region.}
\label{LC_stabilizer}
\end{figure}

First, we will discuss how interaction distances can be reduced in the Louvre-7 circuit, with the example of the $\kappa=2$ La-Cross code~\cite{pecorari_high-rate_2025}. On periodic boundary condition, the La-Cross code can be viewed as a BB code (thus a GB code as well) with generating polynomials $A = 1 +y +y^\kappa$ and $B = 1 + x  + x^\kappa$. The shape of the stabilizers for the $\kappa=2$ La-Cross code is shown in Fig.~\ref{LC_stabilizer}.

We apply the technique discussed in Section~\ref{ordering} to data qubits corresponding to terms in $A$ of Z-ancilla, by arranging the two-qubit interactions in a certain order, and replacing some $\CNOT$ gates with $\CXSWAP$ gates. Meanwhile, we want to preserve the previous reduction in the number of couplers required. First, we construct two gate instructions for Z-ancilla interacting with terms in $A$ and in $B$ separately. Assuming the initial configuration in Fig.~\ref{LC_stabilizer}, we should obtain the gate instructions as follows:

\begin{align}
    & \{A_1\} \to A_2(\CXSWAP)\to \{A_3\} \label{Aorder}\\
    & \{B_1\} \to B_2(\CXSWAP)\to \{B_3\} \label{Border}
\end{align}

For the qubit configuration presented in Fig.~\ref{LC_stabilizer}, the relative positions between data qubits and ancilla qubits are similar to the example discussed in Section~\ref{ordering}, and it is apparent to see that the gate instructions in Eq.~\eqref{Aorder} and Eq.~\eqref{Border} lead to optimal reduction in interaction distances of two-qubit gates. Then, we construct a Louvre-7 circuit by first examining which generating polynomial contains more terms. As before, we assume that $A$ has more terms, and bisect the gate instruction of terms in $A$ in the middle (or near the middle). The first section will be the gate instruction in Phase 1, whereas the second section will be the gate instruction in Phase 3. Ideally, we will use the gate instruction of terms in $B$ as Phase 2 of the syndrome extraction circuit, with a slight notice: By the end of Phase 2, we need to switch the positions of the Z-ancilla and the left data qubits in the basic unit. As displayed in Fig.~\ref{L7BU}, during each routing layer, the Z-ancilla moves from the top right corner to the top left corner in the basic unit, or the other way around. Thus, to have the Z-ancilla at the top left corner of the basic unit in the end of Phase 2 (the position of the left data qubit in the beginning of Phase 2), we require that the number of $\CXSWAP$ routing layers for the Z-ancilla must be odd in Phase 2.  This condition could easily be met, since we can always switch the last layer of Phase 2 from a non-routing layer to a routing layer or vice versa, without affecting the qubit configurations in the rest of Phase 2.

Now we have the gate instruction for Z-ancilla at hand, we can construct the gate instruction for X-ancilla. Viewing Phase 2 as a whole, we need to preserve the relative positions between the X(Z)-ancilla sublattice and the left (right) data qubit sublattice. However, the overall motion of a certain sublattice depends not only on the positions of data qubits with which it interacts via $\CXSWAP$ gate but also on the order in which the interactions happen.

For example, in the La-Cross code example, we have $v_{B_1,Z} = (-1,0)$ and $v_{B_2,Z} = (1,0)$, so if each Z-ancilla $Z_j$ performs the $\CXSWAP$ gate with $q_{B_1,Z_j}$, and then with $q_{B_2,Z_j}$, the movement of the Z-ancilla sublattice will be $(-1,0)$ in the first $\CXSWAP$ gate plus $(1,0) - 2\times(-1,0) = (3,0)$ in the second $\CXSWAP$ gate. The overall movement will be $(2,0)$ viewing Phase 2 as a whole. However, if each X-ancilla $X_j$ performs the $\CXSWAP$ gate with $q_{B_2,X_j}$, and then with $q_{B_1,X_j}$, the overall movement of the right data qubit sublattice will be $(-2,0)$. So the relative positions between the Z-ancilla sublattice and the right data qubit sublattice are not preserved: the right data qubit that used to be at $(0,1)$ of a Z-ancilla is now at $(-4,1)$ of the same ancilla, thus gates in Phase 1 and Phase 3 can no longer share their couplers.

Therefore, we need the routing layers of X-ancilla and Z-ancilla in Phase 2 to interact with data qubits corresponding to the same terms in $B$, in the same order. The easiest way to ensure this is to set the gate instruction for X-ancilla in Phase 2 to be the same as that of Z-ancilla.

Afterward, we set the gate instruction of X-ancilla in Phase 1 (3) to be the same as that of Z-ancilla in Phase 3 (1), but in reversed order, so that we ensure that X-ancilla interact with all their $A_j\in F_x$($F_z$) during Phase 1 (3) using the shared couplers.

\begin{figure}[h!]
\centering
\subfloat[layer 0 (init)]{\includegraphics[width=  0.2\columnwidth]{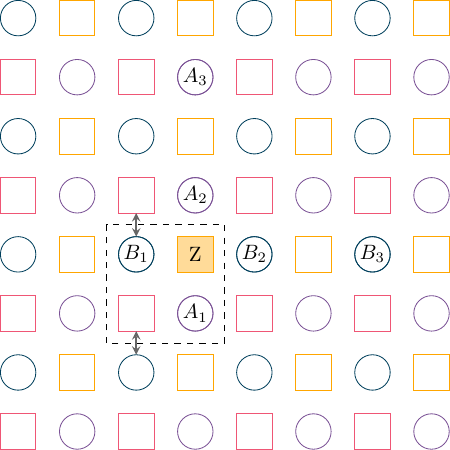}} \hspace{3mm} 
\subfloat[layer 1]{\includegraphics[width=  0.2\columnwidth]{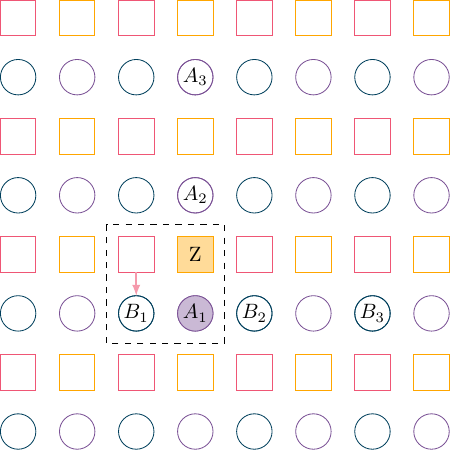}} \hspace{3mm} 
\subfloat[layer 2]{\includegraphics[width=  0.2\columnwidth]{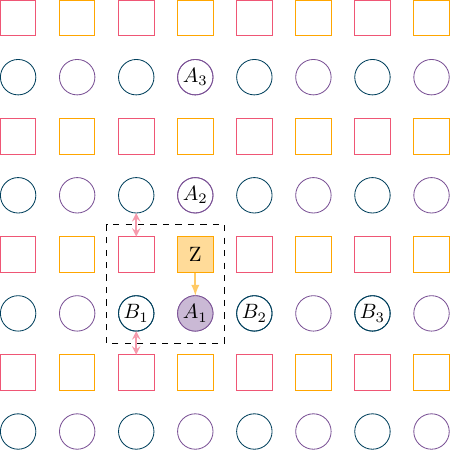}} \hspace{3mm} 
\subfloat[layer 3]{\includegraphics[width=  0.2\columnwidth]{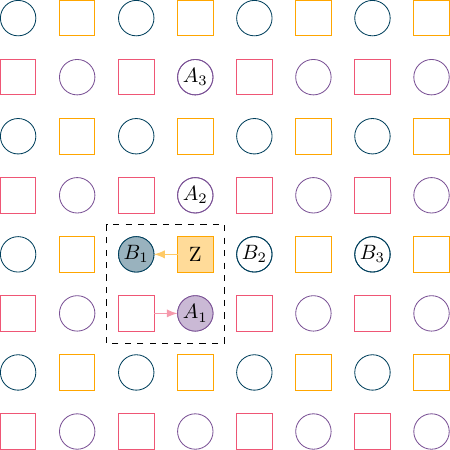}} \\
\subfloat[layer 4]{\includegraphics[width=  0.2\columnwidth]{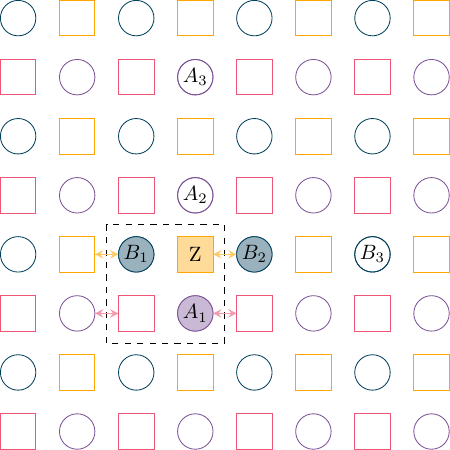}} \hspace{3mm} 
\subfloat[layer 5]{\includegraphics[width=  0.2\columnwidth]{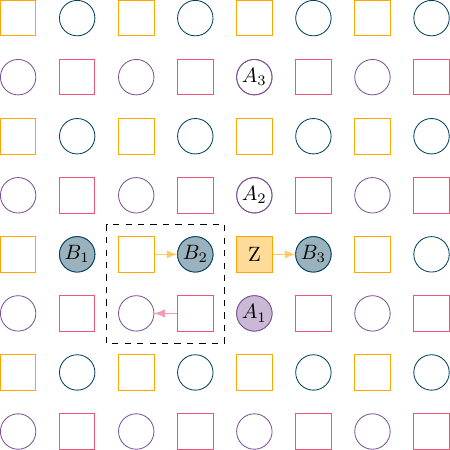}} \hspace{3mm} 
\subfloat[layer 6]{\includegraphics[width=  0.2\columnwidth]{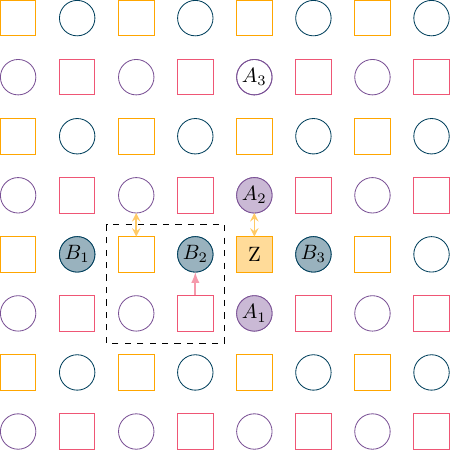}} \hspace{3mm} 
\subfloat[layer 7]{\includegraphics[width=  0.2\columnwidth]{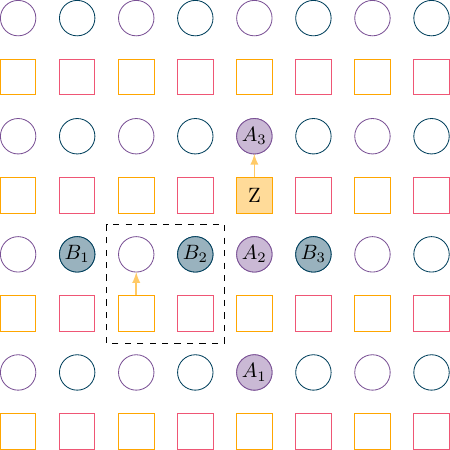}} \\
\subfloat[Overall connectivity required]{\includegraphics[width=  0.4\columnwidth]{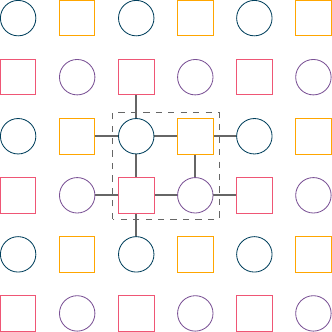}} \\
\caption{(a-h) The  Louvre-7R syndrome extraction circuit for the La-Cross code with interaction distances reduced. (i) The overall connectivity required. Note that the $\kappa=2$ La-Cross code, after reducing interaction distances, can be performed on a hardware with grid topology.}
\label{LC_couplers}
\end{figure}

Finally, we fix the loose end by altering the initial qubit configurations via some \textit{fictional} $\SWAP$ gates to compensate for the $\CXSWAP$ gates in Phase 1. These SWAP gates are fictional because their purposes are to keep track of the changes in the initial qubit configurations, so they do not need to be physically performed. The resulting gate instruction for the $\kappa=2$ La-Cross code is shown in Table~\ref{LCorder}, with the circuit and connectivity requirement shown in Fig.~\ref{LC_couplers}.

\begin{table}[h]
\small
\begin{center}
\caption{The Louvre-7R syndrome extraction circuit of the La-Cross code with reduced interaction distances. Note that a fictional $\SWAP$ need to be performed between all X-ancilla and their corresponding $A_2$ during initialization.}
\label{LCorder}
\begin{tabular}{ |c|c|c|c|c|c|c|c| } 
 \hline
 Phase & \multicolumn{2}{|c|}{1} & \multicolumn{3}{|c|}{2}  & \multicolumn{2}{|c|}{3} \\ 
\hline
Layer & 1&2&3&4&5&6&7 \\ 
\hline
X-ancilla &  $A_3$ & $A_2$($\CXSWAP$) & $B_1$ & $B_2$($\CXSWAP$) & $B_3$ & $A_1$ & \\ 
\hline
Z-ancilla &   & $A_1$ & $B_1$ & $B_2$($\CXSWAP$) & $B_3$ & $A_2$($\CXSWAP$) & $A_3$     \\ 
 \hline
\end{tabular}
\end{center}
\end{table}

\subsection{Louvre-8R}

The application of the same technique in the Louvre-8 circuit has proven to be more complicated, because the Louvre-8 circuit already imposes a condition on the gate instruction during Phase 2 of the syndrome extraction circuit, namely Phase 2A and Phase 2B should be separated by a layer of $\SWAP$ gates. But it is not impossible. Following the discussion of reducing the interaction distances in the Louvre-7R circuit, we have regulated the gate instruction of X- and Z-ancilla during Phase 2 to be copies of each other, consisting of multiple non-routing sections separated by $\CXSWAP$ routing layers. 

As it turns out, one layer of $\SWAP$ gates between all ancilla and their data qubits $A_1$ can reduce the number of couplers required by a non-routing section (or a $\CXSWAP$ routing layer) by half. 

For non-routing sections, the method is similar to the regular Louvre-8 circuit. We divide all the terms within that section into $G_x$ and $G_z$, thus converting the non-routing section into a minor Phase A and Phase B, separated by a layer of $\SWAP$ gates. 

The same could be done for a $\CXSWAP$ routing layer: all Z-ancilla perform $\CXSWAP$ gates with their $B_i$, a layer of $\SWAP$ gates, and then all X-ancilla perform $\CXSWAP$ gates with their $B_i$. Except during the layer of $\SWAP$ gates, ancilla qubits swap with ancilla while data qubits swap with data qubits.

Again, we will need to adjust the initial qubit configuration based on the routing layers in Phase 1. In addition, the modification in the first layer of Phase 3 is only needed when the number of $\SWAP$ routing layers in Phase 2 is odd.

This method imposes a trade-off between the number of couplers, the distances of interaction, and the noise overhead. Each layer of $\SWAP$ gates causes an overhead in gate noise, thus increasing the logical error rate of the code. By reducing the number of $\CXSWAP$ routing layers, we can no longer optimally reduce the distance of the interaction, yet fewer $\CXSWAP$ routing layers lead to more terms in each non-routing section, so that a larger reduction in the number of couplers can be achieved with fewer layers of $\SWAP$ gates. An example of a Louvre-8R circuit for the $\llbracket72,12,6\rrbracket$ BB code with reduced interaction distances is shown in Table~\ref{L8orderred}.

\begin{table}[h]
\small
\begin{center}
\caption{The Louvre-8R syndrome extraction circuit the $\llbracket72,12,6\rrbracket$ BB code with reduced interaction distances.}
\label{L8orderred}
\begin{tabular}{ |c|c|c|c|c|c|c|c|c|c| } 
 \hline
 Phase & \multicolumn{2}{|c|}{1} & \multicolumn{5}{|c|}{2}  & \multicolumn{2}{|c|}{3} \\ 
\hline
Layer & 1&2&3&4&5&6&7&8&9 \\ 
\hline
X-ancilla &  $A_3$ & $A_2$($\CXSWAP$) & $B_1$ & $B_2$($\CXSWAP$) & $B_3$ & $A_2$($\SWAP$)& & $A_2$($\SWAP$)&$A_1$ \\ 
\hline
Z-ancilla &   & $A_1$ & $B_1$ & $B_2$($\CXSWAP$) & & $A_2$($\SWAP$)&$B_3$ & $A_2$ & $A_3$     \\ 
 \hline
\end{tabular}
\end{center}
\end{table}

\subsection{Special Case: $\CXSWAP$ Only}

It is interesting to note that, though the Louvre-7 circuit is developed under the assumption that we have multiple two-qubit gates in our native instruction set, for the $\kappa=2$ La-Cross code, as well as the BB code, the circuit can be performed with only $\CXSWAP$ gates: replacing the rest of $\CNOT$ gates in Table~\ref{LCorder} into $\CXSWAP$ gates will only alter the initial configuration of the code, without affecting its function or performance. It merely imposes an additional condition when compiling the gate instructions, that all the non-routing sections are empty sets, and all gates are performed during routing layers. The same can be done on Louvre-7 circuits for BB codes, though the additional condition on gate instruction might cause the distance of some interaction to increase rather than decrease.

Attempting the same thing on Louvre-8 circuits requires some slight complication, but it can be done nevertheless. From the circuit in Table~\ref{L8orderred}, we will be able to obtain a circuit that contains both $\CXSWAP$ gates and $\SWAP$ gates, with some additional overhead in noise and circuit depth. One example is shown in Table~\ref{L8alliswap}, where the circuit is of depth-11 and 2 layers of $\SWAP$ gates are inserted.

\begin{table}[h]
\small
\begin{center}
\caption{An alternative Louvre-8R syndrome extraction circuit of the La-Cross code. Note that the default two-qubit gates is $\CXSWAP$ unless otherwise specified.}
\label{L8alliswap}
\begin{tabular}{ |c|c|c|c|c|c|c|c|c|c|c| } 
 \hline
 Phase & \multicolumn{2}{|c|}{1} & \multicolumn{6}{|c|}{2}  & \multicolumn{2}{|c|}{3} \\ 
\hline
Layer & 1&2&3&4&5&6&7&8&9&10 \\ 
\hline
X-ancilla &  $A_3$ & $A_2$ & $B_1$ & $B_2$ & $A_2$($\SWAP$)& $B_3$ & $A_2$($\SWAP$)& &$A_1$ & \\ 
\hline
Z-ancilla &   & $A_1$ & $B_1$ & &$A_2$($\SWAP$)& $B_2$ & $A_2$($\SWAP$)&$B_3$ & $A_2$ & $A_3$     \\ 
 \hline
\end{tabular}
\end{center}
\end{table}

\section{Results}
\label{sec:numerics}

In this section, we present numerical results about Louvre circuits. We will show how their logical error rates compare to the regular circuit, as well as the effects of the reduction of the coupler length between the superconducting qubits under realistic metrics. The relevant parameters of some representative codes are listed in Table~\ref{code_parameter}. Some additional La-Cross codes are listed in Table~\ref{LCparameters}.

\begin{table}[h]
\small
\begin{center}
\caption{The parameters of the codes discussed in this paper.}
\label{code_parameter}
\begin{tabular}{ |c|c|c|c|c|c|c| } 
 \hline
  $\llbracket n,k,d\rrbracket$&l&m& $A$ & $B$ & Code Type & Source \\ 
\hline
\hline
$\llbracket18,4,4\rrbracket$& 3&3& $1+y+xy$ & $1+x+xy$ & BB code & \cite{wang_demonstration_2025}\\ 
\hline
$\llbracket72,12,6\rrbracket$& 6&6& $y+y^2+x^3$ & $y^3+x+x^2$ & BB code & \cite{bravyi_high-threshold_2024}\\ 
\hline
$\llbracket108,8,10\rrbracket$& 9&6& $y+y^2+x^3$ & $y^3+x+x^2$ & BB code & \cite{bravyi_high-threshold_2024}\\ 
\hline
$\llbracket72,8,4\rrbracket$& 6&6& $1+y+y^2$ & $1+x+x^2$ & La-Cross code & \cite{pecorari_high-rate_2025}\\ 
\hline
$\llbracket72,8,9\rrbracket$&9&4& $1+y$ & $1+x+y^{6}+x^{3}y+xy^{7}+x^{3}y^5$ & GB code & \cite{lin_quantum_2023}\\ 
\hline 
$\llbracket72,8,10\rrbracket$& 36&1& $1+x^{9}+x^{28}+x^{31}$ & $1+x+x^{21}+x^{34}$ & GB code & \cite{lin_quantum_2023}\\ 
\hline
$\llbracket96,10,12\rrbracket$& 12&4& $1+y+xy+x^9$ & $1+x^2+x^7+x^9y^2$ & GB code & \cite{lin_quantum_2023}\\ 
\hline
$\llbracket128,16,8\rrbracket$& 8&8& $y+y^2+y^5+x^6$ & $y^2+x^2+x^3+x^7$ & GB code & \cite{viszlai_matching_2024}\\ 
\hline
\end{tabular}
\end{center}
\end{table}

\subsection{Logical Error Rate}
We have simulated the logical error rate of numerous different quantum error correcting codes, including BB code, La-Cross code, and GB code, under the SI(1000) noise model \cite{Gidney2022benchmarkingplanar}. The logical error rates presented here are world logical errors, where we consider an error to have occurred when an error occurs on any of the logical qubits. We compare their logical error rate over 6 rounds with syndrome extraction circuits:
\begin{enumerate}
    \item Static circuit (Regular)
    \item Louvre-7 without interaction distances reduction (Louvre-7)
    \item Louvre-7 with interaction distances reduction (Louvre-7R)
    \item Louvre-8 without interaction distances reduction (Louvre-8)
    \item Louvre-8 with interaction distances reduction (Louvre-8R)
\end{enumerate}
Note that not all five types of circuits are simulated for every code, because for certain code, some Louvre circuits do not lead to apparent improvement. For example, Louvre-7R does not further reduce the interaction distance in the $\llbracket18,4,4\rrbracket$ BB code compared to the regular Louvre-7 circuit. In some other case, it seems unnecessary to consider interaction distances without a reasonable qubit configuration.

\subsubsection{Bivariate Bicycle Codes}
\begin{figure}[h]
\centering
\subfloat[$\llbracket 18,4,4\rrbracket$ BB code]{\includegraphics[width=  0.3\columnwidth]{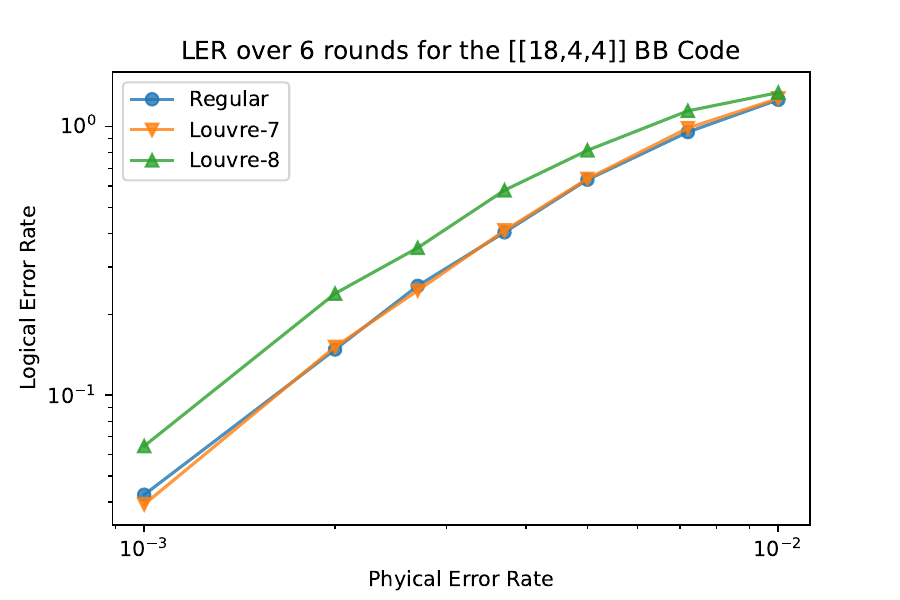}} \hspace{5mm}
\subfloat[$\llbracket 72,12,6\rrbracket$ BB code]{\includegraphics[width=  0.3\columnwidth]{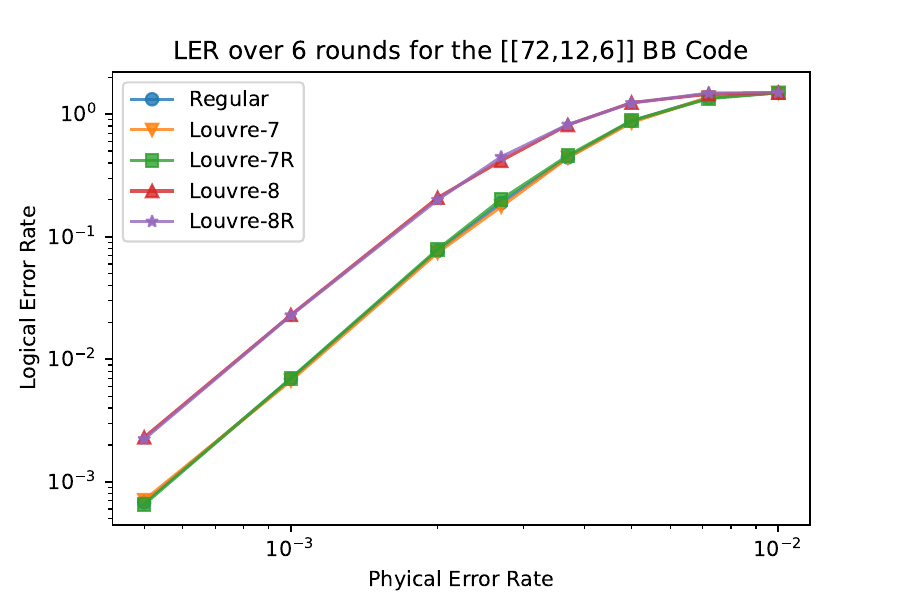}} \hspace{5mm}
\subfloat[$\llbracket 108,8,10\rrbracket$ BB code]{\includegraphics[width=  0.3\columnwidth]{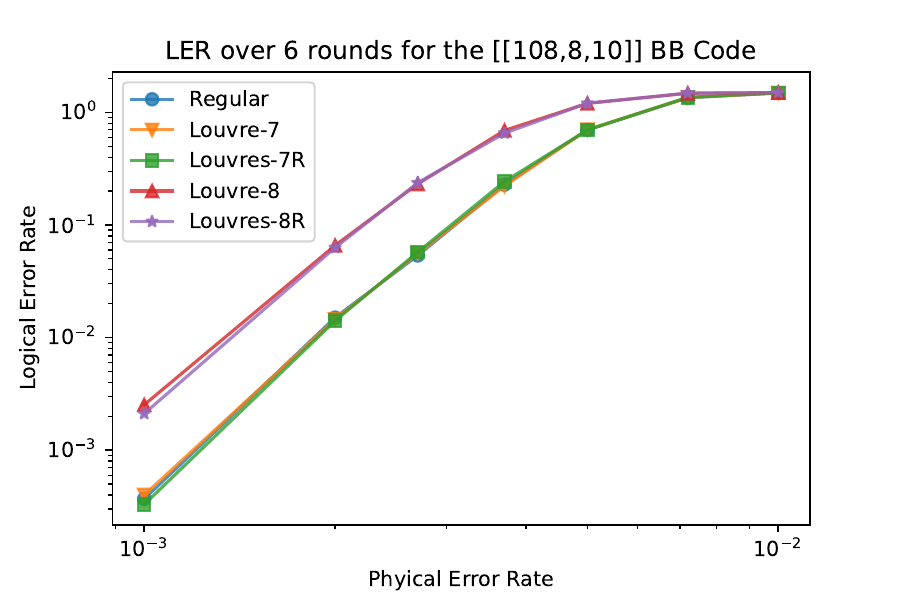}}
\caption{The logical error rate  with different syndrome extraction circuit. The logical error rate is the world logical error rate, that is, we consider a error occur when a error occur in any of the logical qubits.}
\label{BB_LER}
\end{figure}

We have simulated the logical error rate of some BB codes, including the $\llbracket18,4,4\rrbracket$ BB code introduced as an example in Section~\ref{reduce_number}. Without distance reduction, the longest interaction for both $\llbracket 72,12,6\rrbracket$ and $\llbracket 108,8,10\rrbracket$ BB code have distance of 9 in terms of L1 distance, which can be reduced to 7 using Louvre-7R and Louvre-8R circuits. The result is shown in Fig.~\ref{BB_LER}. We see that the Louvre-7 and Louvre-7R circuit is equivalent to the static circuit in terms of its logical error rate, while the Louvre-8 circuit endures some additional logical error due to the inserted layer of $\SWAP$ gates. The logical error rate of the Louvre-8R circuit appears to be similar to that of the Louvre-8 circuit, indicating that the effect of the additional idling error is small.

\begin{figure}[h]
\centering
\includegraphics[width=  0.5
\columnwidth]{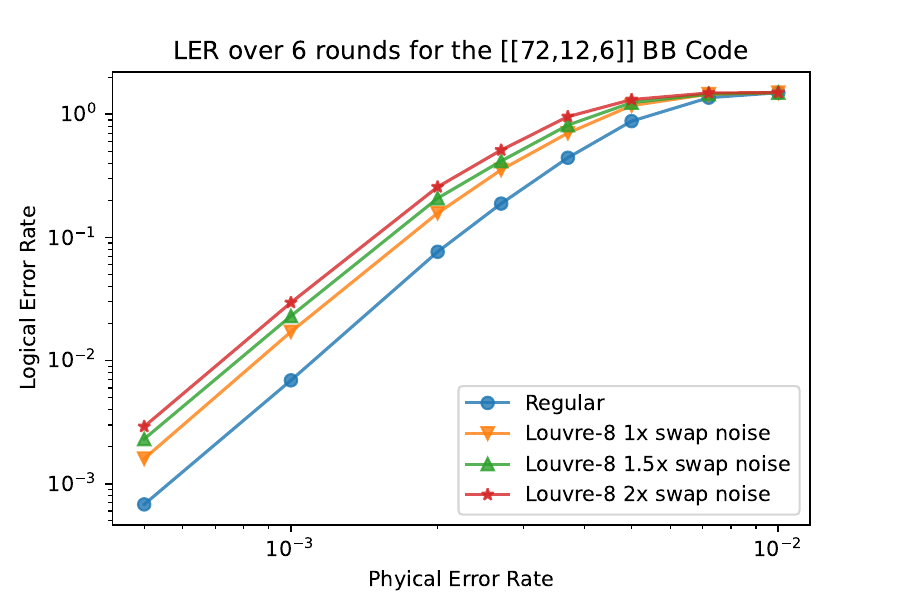} 
\caption{The logical error rate of the $\llbracket 72,12,6\rrbracket$ BB code with different gate nosie factor of the $\SWAP$ gate.}
\label{BB_swap_noise}
\end{figure}

We have also examined the effect of adjusting the gate noise factor for the $\SWAP$ gates on the Louvre-8 circuit, as shown in Fig.~\ref{BB_swap_noise}. As expected, the logical error rate of the code increases as the noise induced by $\SWAP$ increases.

\subsubsection{La-Cross Code}
\begin{figure}[h]
\centering
\includegraphics[width=  0.5
\columnwidth]{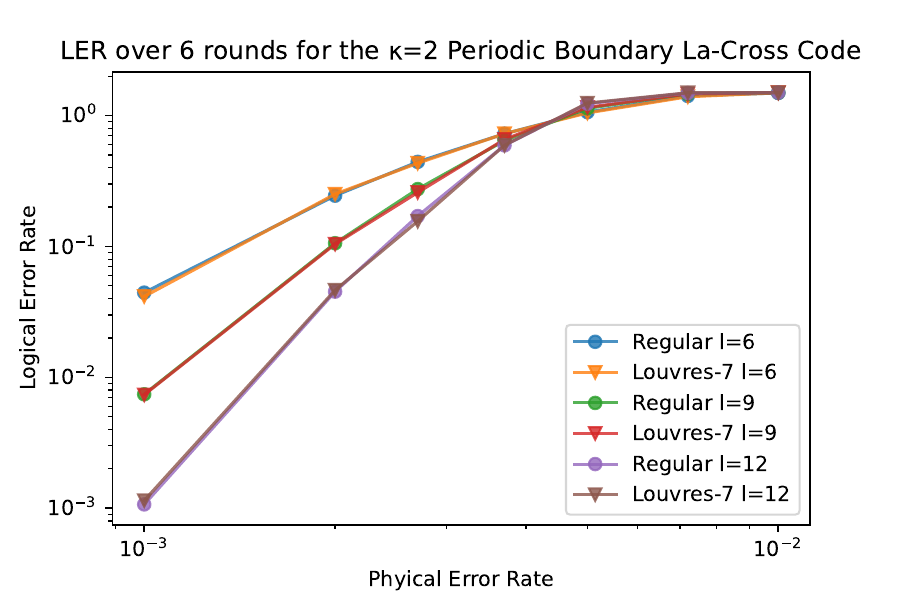} 
\caption{The logical error rate of the $\kappa=2$ La-Cross code on the periodic boundaries with different syndrome extraction circuit.}
\label{PBLC_LER}
\end{figure}

We simulated the $\kappa=2$ La-Cross code on the periodic boundaries with different code size, as introduced in Section~\ref{reduce_length}, and compared their logical error rate over 6 rounds with regular syndrome extraction circuit and Louvre-7R, which can be performed on grid topology. The result is shown in Fig.~\ref{PBLC_LER}. Again, we see that the Louvre-7R circuit is equivalent to the static circuit in terms of its logical error rate.

\begin{figure}[h!]
\centering
\subfloat[ $\kappa=2$ La-Cross code]{\includegraphics[width=  0.45\columnwidth]{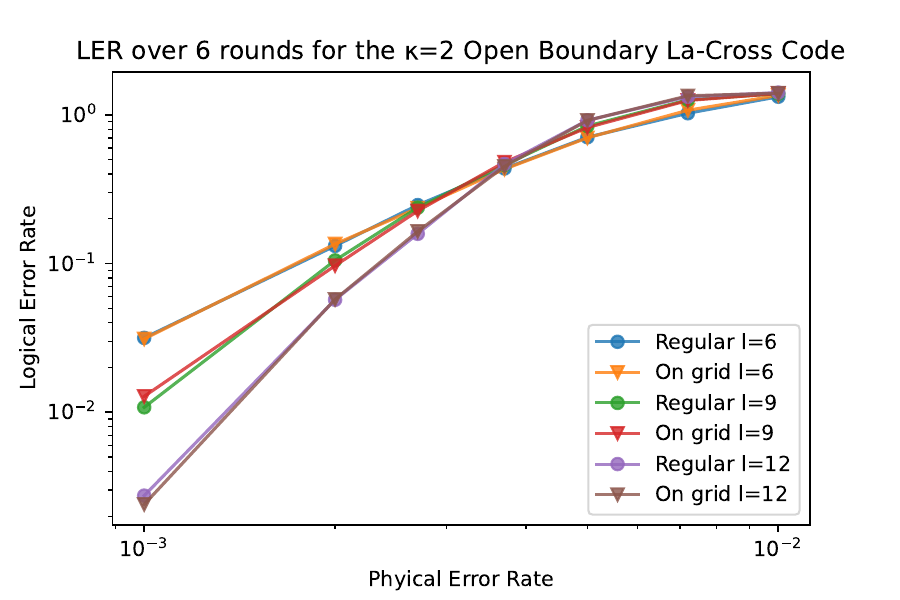}} \hspace{3mm} 
\subfloat[ $\kappa=3$ La-Cross code]{\includegraphics[width=  0.45\columnwidth]{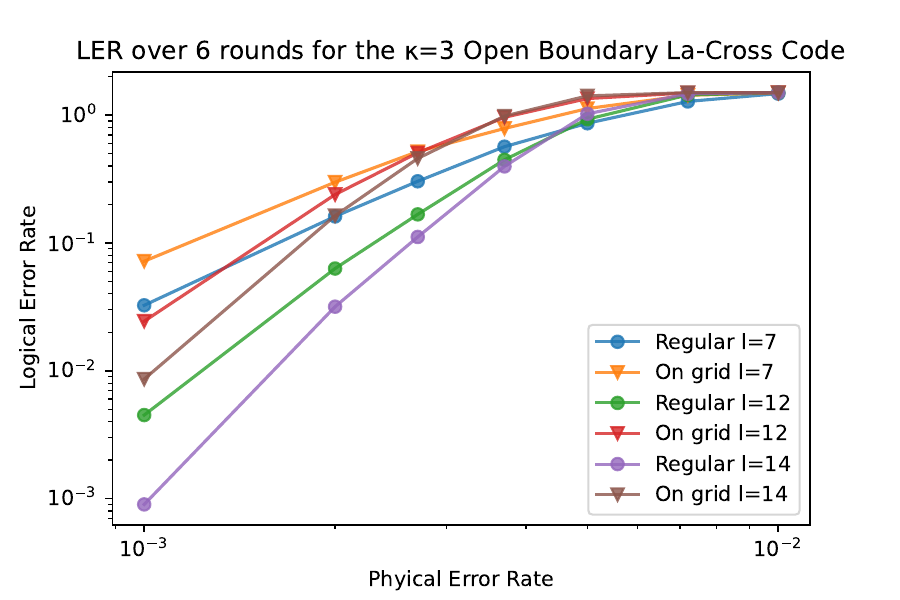}} 
\caption{The logical error rate of the (a) $\kappa=2$ and (b) $\kappa=3$ La-Cross code on open boundaries with different syndrome extraction circuit.}
\label{OBLC_LER}
\end{figure}

To meet the current hardware, we also simulated some La-Cross codes on open boundaries. The conversion can be understood by viewing the La-Cross code as a hypergraph product code. The parameters of the codes simulated are presented in the Table~\ref{LCparameters}. We compare the logical error rate with regular syndrome extraction and with the Louvre-7R circuit. The result is shown in Fig.~\ref{OBLC_LER}. For the $\kappa=2$ La-Cross code, the Louvre-7 circuit maintains an equivalent logical error rate with the regular circuit. However, to the $\kappa=3$ La-Cross code on grid topology, some addition $\SWAP$ gates need to be inserted to facilitate the long-ranged interactions required, therefore the logical error rate is significantly higher.

\begin{table}[h]
\small
\begin{center}
\caption{Relevant parameters of the La-Cross code discussed in this paper. La-Cross codes can be viewed as an BB code with generating polynomials $A$ symmetric to $B$; or as a hypergraph product code, generated by two classical seed codes.}
\label{LCparameters}
\begin{tabular}
    {|c|c|c|c|c|c|}
    \hline
    Generating Polynomial   & Boundary  & Seed code  & Parameters    & Code rate\\
    \hline
    \hline
    $1+x+x^2$   & Periodic  & [6,2,4]   & [[72,8,4]]    & 0.111\\
    $1+x+x^2$   & Periodic  & [9,2,6]   & [[162,8,6]]   & 0.049\\
    $1+x+x^2$   & Periodic  & [12,2,8]  & [[288,8,8]]   & 0.028\\
    \hline
    $1+x+x^2$   & Open      & [6,2,4]   & [[52,4,4]]    & 0.077\\
    $1+x+x^2$   & Open      & [9,2,6]   & [[130,4,6]]   & 0.031\\
    $1+x+x^2$   & Open      & [12,2,8]  & [[244,4,8]]   & 0.016\\
    \hline
    $1+x+x^3$   & Open      & [7,3,4]   & [[65,9,4]]    & 0.138\\
    $1+x+x^3$   & Open      & [12,3,6]  & [[225,9,6]]   & 0.040\\
    $1+x+x^3$& Open      & [14,3,8]  & [[317,9,8]]   & 0.028\\
    \hline
\end{tabular} 
\end{center}
\end{table}

\subsubsection{Generalized Bicycle Codes}
\begin{figure}[h]
\centering
\subfloat[$\llbracket 72,8,9\rrbracket$ GB code]{\includegraphics[width=  0.45\columnwidth]{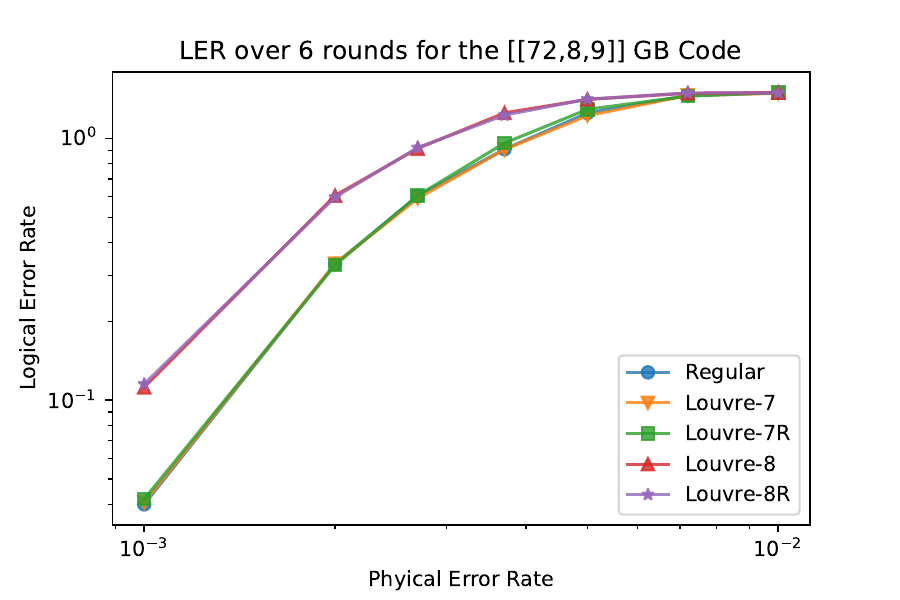}} \hspace{5mm}
\subfloat[$\llbracket 72,8,10\rrbracket$ GB code]{\includegraphics[width=  0.45\columnwidth]{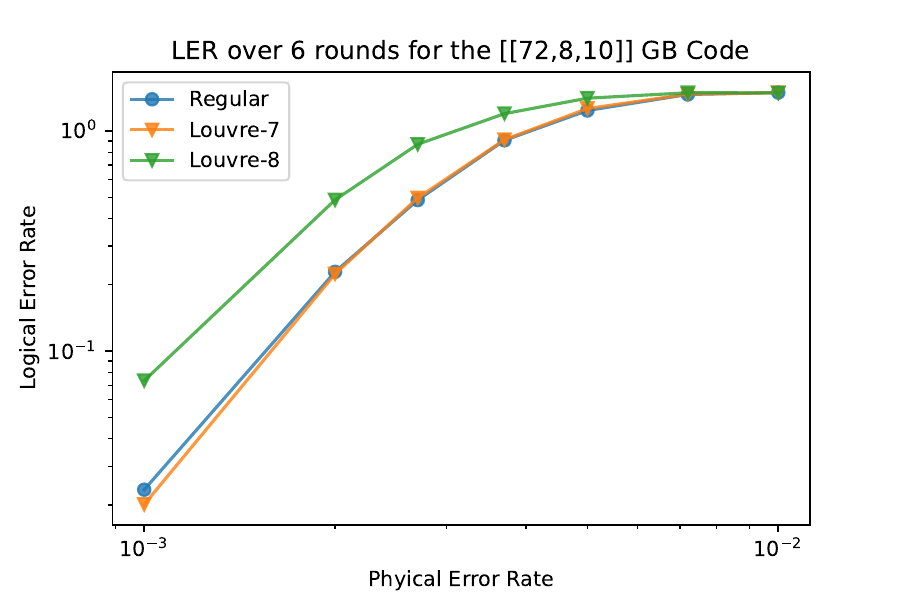}} \\
\subfloat[$\llbracket 96,10,12\rrbracket$ GB code]{\includegraphics[width=  0.45\columnwidth]{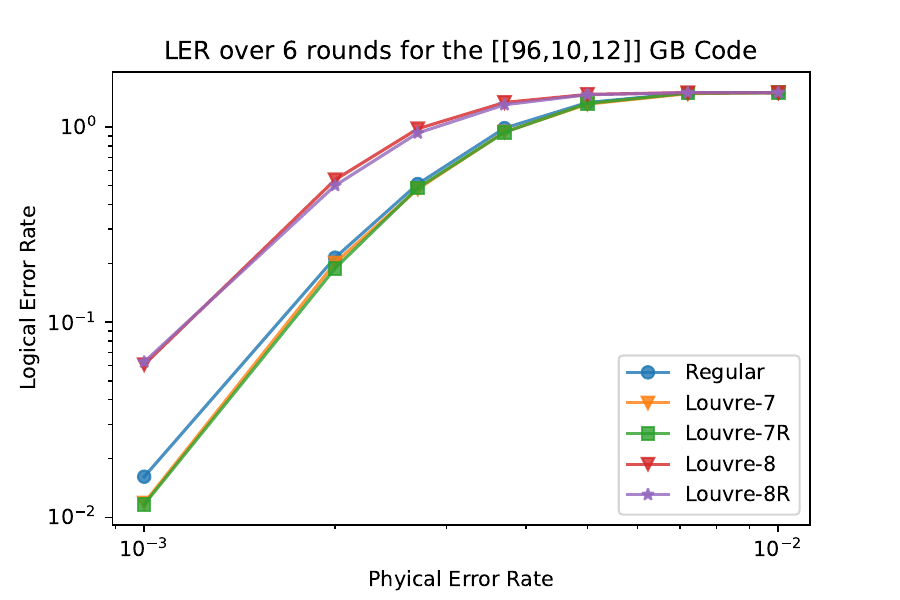}} \hspace{5mm}
\subfloat[$\llbracket 128,16,8\rrbracket$ GB code]{\includegraphics[width=  0.45\columnwidth]{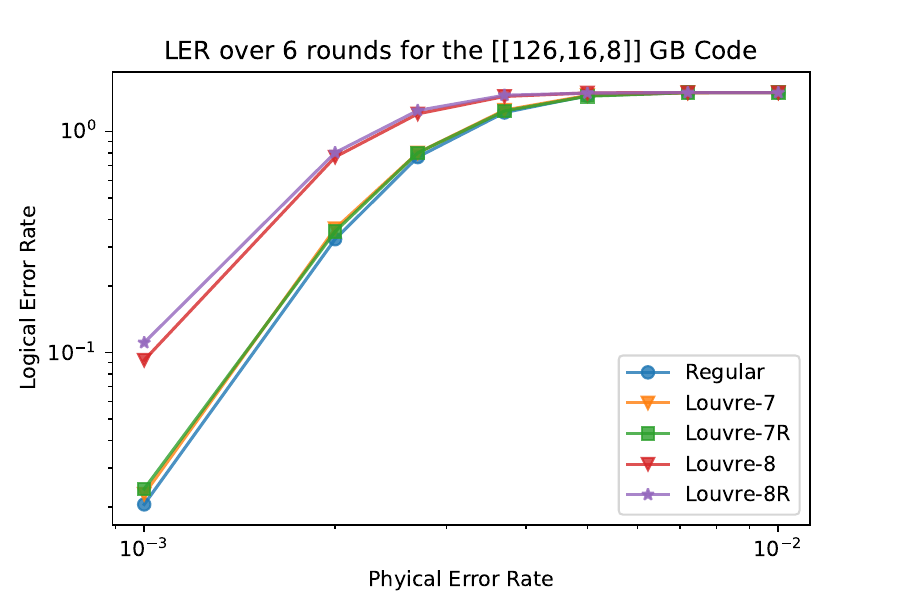}} \\
\caption{The logical error rate of the generalized bicycle code with different syndrome extraction circuit.} 
\label{GB_LER}
\end{figure}

We have simulated the logical error rate of some generalized bicycle codes, under the SI(1000) noise model, and compare their logical error rate over 6 rounds with different syndrome extraction circuits. The result is shown in Fig.~\ref{GB_LER}. 

Note that the $\llbracket72,8,9\rrbracket$ code shown in Fig.~\ref{GB_LER}(a) only has two terms in $A$, both corresponding to short-ranged interactions. Thus, the Louvre-8 and Louvre-8R circuits do not provide significant benefit beyond Louvre-7 and Louvre-7R, as shown in Table~\ref{code_results}.

\subsection{Coupler Length}
\label{subsec:length_numerics}

Although a multi-layer structure employing bump-bonding, TSVs and long-range couplers can enable the connectivity required for error correction codes, their implementation inevitably introduces degradation in the fidelity of qubit operations and increases the fabrication failure rates. Minimizing the number of these components used on the chip will greatly help the implementation of the error correction codes with long-range interactions on superconducting qubits. 

To enable end-to-end code deployment, we have developed placement and routing modules for LDPC codes in multi-layer superconducting hardware, utilizing these modules to analyze how reducing coupler lengths and degrees can mitigate hardware requirements. In preparing this manuscript, we identified a recent study~\cite{mathews2025placingroutingnonlocalquantum} that focuses on hardware requirements for deploying LDPC codes yet lacks the code optimization that constitutes our work’s primary contribution. For the reader’s convenience, we adopt the same performance metrics as employed in~\cite{mathews2025placingroutingnonlocalquantum}. That said, we wish to clarify that while we share certain similar concepts and adopt some configurations from~\cite{mathews2025placingroutingnonlocalquantum}, our implementations and some configurations remain distinct---differences that may introduce discrepancies in results. Importantly, we leverage our in-house placement and routing modules to demonstrate the advantage of relaxing hardware requirements when using Louvre.

In a superconducting quantum processor, chips can be stacked vertically, with each chip providing two surfaces, referred to layers, to place qubits and couplers. A pair of opposing layers from two adjacent chips are defined as a tier that is connected via bump-bonding. Two adjacent tiers are connected by TSVs in the chip between them. In our implementation, the first tier is arranged as a square grid to enable direct coupling, while long-range couplers are allocated to higher tiers. For each subsequent tier, edges are routed sequentially in ascending order of their straight-line lengths. Each edge is processed using an A* pathfinding algorithm on an $[N, M, 2]$ occupancy grid, with a maximum inter-layer movement constraint of 10, which is supposed to be realized through bump-bonding. Upon successful routing, the grid cells along the path are marked as occupied. Edges that cannot be routed in the current tier are deferred to the next tier. All qubits requiring further routing are replicated to the subsequent tier which is supposed to be connected using TSVs. To evaluate hardware cost, we record the number of tiers, the average coupler length, the number of bump-bonding connections, and the number of TSVs. The results can be found in Table~\ref{tab:multi-layer}. Once again, we must stress that the qubit configurations and Louvre circuits used here are constructed from human ansatz, instead of exhaustive algorithmic search. These implementations of the codes are not necessarily optimal, nor are their results from the routing algorithm. Nevertheless, we can see that Louvre circuits can generally reduce the number of tiers by 1-2 in comparison to the regular syndrome extraction circuit for the BB and GB codes simulated. It is also interesting to note that, aside from the number of tiers, the Louvre circuits with interaction distances reduced (Louvre-7R and Louvre-8R) usually lead to further improvement from the original circuits (Louvre-7 and Louvre-8) in terms of the number of TSVs, but not necessarily in coupler lengths and the number of bump-bonding. So, it is not straightforward to conclude whether these schemes further relax the requirement imposed on superconducting hardware. 

\begin{table}[h]
\small
\begin{center}
\caption{Hardware cost on a multi-layer superconducting device. "Tiers" column is  the number of tiers, "Length" column is the average long-range coupler length in the unit of shortest distance between qubits, "Bumps" column is the average number of bump-bonding per long-range coupler, "TSVs" column is the average number of TSVs per long-range coupler.}
\label{tab:multi-layer}
\begin{tabular}{ |c|c|c|c|c|c|c| } 
 \hline
  $\llbracket n,k,d\rrbracket$ & Code Type & Scheme & Tiers & Length & Bumps & TSVs \\ 
\hline
\hline
\multirow{3}{*}{$\llbracket18,4,4\rrbracket$} & \multirow{3}{*}{BB Code} & Regular &3& 4.11 & 2.04 & 1 \\
 & & Louvre-7 & 2 & 4.18 & 1.78 & 1 \\
 & & Louvre-8 & 2 & 3.84 & 1.25 & 1.33 \\
\hline
\multirow{5}{*}{$\llbracket72,12,6\rrbracket$} & \multirow{5}{*}{BB Code}  & Regular & 4 & 9.25 & 4.71 & 1.74\\
 &  & Louvre-7 & 4 & 8.2 & 3.9 & 1.74\\
 &  & Louvre-7R & 3 & 8.42 & 3.43 & 1.6\\
 &  & Louvre-8 & 3 & 7.83 & 3.4 & 1.86\\
 &   & Louvre-8R & 3 & 8.66 & 3.47 & 1.62\\
\hline
\multirow{5}{*}{$\llbracket144,12,12\rrbracket$} & \multirow{5}{*}{BB Code}  &  Regular & 4 & 10.36 & 4.04 & 2.01\\
 &  & Louvre-7 & 4 & 9.62 & 3.84 & 1.89\\
 &  & Louvre-7R & 3 & 9.71 & 3.09 & 1.77\\
 &  & Louvre-8 & 3 & 9.64 & 3.3 & 1.88\\
 &  & Louvre-8R & 3 & 10.39 & 3.36 & 1.74\\
\hline
\multirow{5}{*}{$\llbracket72,8,9\rrbracket$} & \multirow{5}{*}{GB Code}  & Regular & 5 & 9.16 & 3.16 & 2.06\\
 &  & Louvre-7 & 3 & 8.3 & 3.22 & 1.68\\
 &  & Louvre-7R & 3 & 8.46 & 3.39 & 1.58\\
 &  & Louvre-8 & 3 & 8 & 3.1 & 1.74\\
 &  & Louvre-8R & 3 & 8.1 & 3.16 & 1.7\\
 \hline
\multirow{4}{*}{$\llbracket96,10,12\rrbracket$} & \multirow{4}{*}{GB Code}  & Regular & 5 & 10.43 & 2.92 & 2.24\\
 &  & Louvre-7 & 5 & 9.14 & 2.61 & 1.67\\
 &  & Louvre-7R & 4 & 8.71 & 2.39 & 1.76\\
 &  & Louvre-8 & 4 & 9.31 & 2.96 & 1.7\\
 &  & Louvre-8R & 4 & 8.69 & 2.34 & 1.69\\
 \hline
\multirow{5}{*}{$\llbracket128,16,8\rrbracket$} & \multirow{5}{*}{GB Code}  & Regular & 5 & 9.75 & 4.14 & 2.33\\
 &  & Louvre-7 & 5 & 8.89 & 4 & 2.23\\
 &  & Louvre-7R & 4 & 8.99 & 3.19 & 1.92\\
 &  & Louvre-8 & 4 & 8.73 & 3.82 & 2.08\\
 &  & Louvre-8R & 4 & 8.89 & 3.11 & 1.91\\
 \hline
\end{tabular}
\end{center}
\end{table}

We have also studied the placing problem in another setting, based on the consideration that the main advantage of BB codes is their thickness-2 connections. By placing the qubits and some of the couplers on a 2D grid, it is known that the remaining couplers form a planar graph and could be embedded into the second layer without crossing each other. As the locations of the qubits are fixed due to geometry on the first layer, the placement of edges on the second layer could be abstracted into the problem of embedding planar graphs with fixed vertex locations. Finding the minimum-length embedding was shown to be NP-hard \cite{bastert1998geometric}, and several algorithms have been proposed to generate an approximate solution \cite{pach2001embedding,chan2013minimum,schaefer2021new}. Here, we implemented the algorithm in \cite[Thm.~11]{schaefer2021new} with the spanning forest constructed from \cite[Lemma 5]{chan2013minimum}. To our knowledge, this has the best bound on the solution among all the algorithms provided in these papers. Note that there are some arbitrariness in the algorithm, e.g., the order of the edges when generating the spanning forest. In the implementation of the algorithm, we make these choices randomly and find the best solution among multiple runs. We apply the algorithm to various BB codes and list the results in Table~\ref{tab:two-layer-embedding}. Given the computational intractability of the exact minimum-length embedding, we use the sum of the straight-line distance of the edges as a baseline, and study the ratio of the solution's total path length to this baseline value. As we can see, by removing some of the edges, the Louvre schemes have not only a smaller total length than the original scheme but also a smaller ratio to the baseline. This is consistent with the intuition that routing becomes easier with fewer edges. However, the advantages of Louvre-7R and 8R over Louvre-7 and 8 are not conclusive from the results, as the ratio slightly increases despite the decrease in the total routed length.

\begin{table}[ht]
    \caption{Hardware cost on a two-layer superconducting device.}
    \label{tab:two-layer-embedding}
    \centering
\begin{tabular}{ |c|c|c|c|c| } 
 \hline
  $\llbracket n,k,d\rrbracket$ & Scheme & Straight-line Distance & Routed Distance & Ratio \\ 
\hline
\hline
\multirow{3}{*}{$\llbracket18,4,4\rrbracket$} & Regular & 120.2993 & 1254.3732 & 10.4271 \\
 & Louvre-7 & 94.5136 & 951.6118 & 10.0685 \\
 & Louvre-8 & 60.1497 & 397.5847 & 6.6099 \\
\hline
\multirow{6}{*}{$\llbracket72,12,6\rrbracket$} & Regular & 1113.8856 & 33501.8452 & 30.0766 \\
  & Louvre-7 & 847.7395 & 24957.7053 & 29.4403 \\
  & Louvre-7R & 695.6216 & 26223.0593 & 37.6973 \\
  & Louvre-8 & 556.9428 & 9693.7013 & 17.4052 \\
  & Louvre-8R & 476.6421 & 9980.0883 & 20.9383 \\
\hline
\multirow{6}{*}{$\llbracket144,12,12\rrbracket$} &  Regular & 2795.7868 & 115134.8405 & 41.1816\\
  & Louvre-7 & 2062.4696 & 83751.8786 & 40.6076 \\
  & Louvre-7R & 1596.1132 & 87742.0639 & 54.9723\\
  & Louvre-8 & 1397.8934 & 39365.1660 & 28.1603 \\
  & Louvre-8R & 1158.1543 & 35549.0352 & 30.6946 \\
 \hline
\end{tabular}\end{table}

\section{Discussion}
\label{sec:discussion}

In developing the method of Louvre, we took a top-down approach, attempting to relax the requirement imposed on hardware connectivity when a generalized bicycle code is given. Our result has demonstrated that the Louvre-7 and Louvre-7R circuits are equivalent to the regular syndrome extraction circuit in terms of their logical error rate. And the Louvre-8 and Louvre-8R circuits will induce some noise overhead caused by the inserted $\SWAP$ gates, causing the logical error rate to be typically $\sim3\times$ that of the regular circuit.

The method of Louvre is developed under the assumption of periodic boundary conditions and that the grid is complete (a qubit exists as part of the code on every grid point). Nevertheless, as shown in the Appendix~\ref{missing_qubit}, Louvre can be adjusted to codes with open boundaries or when some qubits are missing in the grid. In addition, as in other recent research discussing the implementation of BB codes on superconducting hardware, we have assumed that interaction between distanced qubits can be facilitated via long-ranged couplers. Although this has been demonstrated in recent hardware experiments~\cite{wang_demonstration_2025}, some of the codes in our discussion, namely the La-Cross code, can be performed with only nearest-neighbor interactions, the hardware for which has been readily available in surface code experiments~\cite{acharya2024quantum}. Meanwhile, we have also assumed that our native instruction set contains multiple types of two-qubit gates, including the $\CNOT$ gate, the $\CXSWAP$ gate for Louvre-7, and in addition the $\SWAP$ gate for Louvre-8. However, for some GB codes, including BB codes and La-Cross code, the Louvre-7 circuit can be performed with only native $\CXSWAP$ gates. All of these features make our results immediately relevant to the status of current hardware experiments.

\section*{Acknowledgements}
This work is supported by Zhongguancun Laboratory. J.C. would like to thank Dawei Ding for numerous insightful discussions on the potential advantages of expanding native gate sets in QECC. Though this work has not yet fully leveraged the AshN scheme~\cite{chen_one_2024,chen2025efficientimplementationarbitrarytwoqubit}, it nonetheless suffices to highlight its potential.

\appendix

\section{Adaption to Open Boundaries and Grids with Absent Sites} 
\label{missing_qubit}

So far our discussion has been focusing on the generalized bicycle code with periodic boundary condition, where all the qubits in all the basic units are active parts of the code. However, when considering experimental implementation, this might not be the case. In some experiment, a small part of ancilla qubits are absent on the hardware because their corresponding stabilizers are redundant \cite{wang_demonstration_2025}. In another case, when implementing BB codes on an open boundary hardware \cite{liang_planar_2025}, some stabilizers on the boundary of the code do not include data qubits associated with all the terms in the generating polynomial. 

These situations affect Lourvre in that, during a routing layer, where all ancilla $X_j$ are supposed to swap their positions with $q_{B_1,X_j}$ for example, it could be the case that some ancilla qubits $X_j$($Z_j$) or data qubits $q_{B_1,X_j}$ are absent in code, so the relevant routing cannot be performed. When these situations happen, we call the absent qubit an \textit{absent site}. When absent sites exist in the code, they breaks the premise that all qubits in a sublattice move together in a routing layer, thus complicates things when considering relative positions between qubits. Note that an absent site differs from a hardware defect, in that it is not caused by incidental hardware imperfection, but rather a man-made choice when designing or performing a QECC.

The resolution of this circumstance is straightforward. Again we will walk through it with the Louvre-7 circuit of the $\llbracket 18,4,4\rrbracket$ BB code as an example. We consider two different scenarios. In the first scenario, we assume that the hardware is large and parallel in its connectivity. That is, the hardware contains a plentiful amount of basic units in the encoded region and its surroundings, and the connectivity in each basic unit is the same. In this case, although the absent site is not part of the code, there should be a qubit in the corresponding position, and we call that a padding qubit. We could manually perform a $\SWAP$ gate (or an equivalent $\iSWAP$ gate, by resetting the padding qubit at the $\ket{0}$ or $\ket{+}$ state) between the encoded qubit and the padding qubit, so that the premise regarding sublattice moving together still holds, and the following circuit can be carried out. The additional $\SWAP$ gate will introduce noise and affect the resulting logical error rate, but the effect should be mild.

\begin{figure}[h]
\centering
\subfloat[]{\includegraphics[width=  0.25\columnwidth]{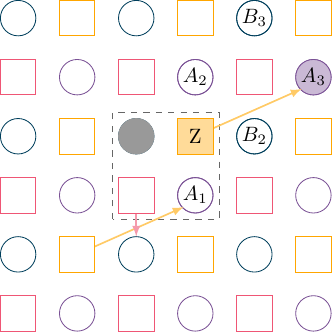}} \hspace{5mm} 
\subfloat[]{\includegraphics[width=  0.25\columnwidth]{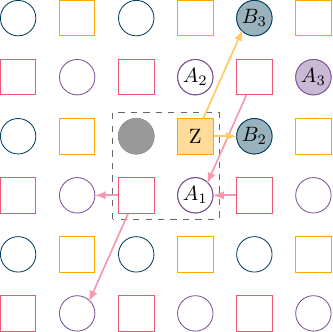}} \hspace{5mm} 
\subfloat[]{\includegraphics[width=  0.25\columnwidth]{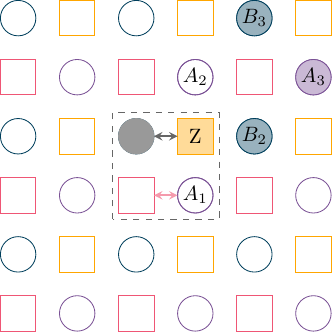}} \\
\subfloat[]{\includegraphics[width=  0.25\columnwidth]{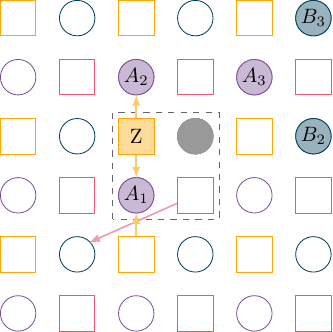}} \hspace{5mm} 
\subfloat[]{\includegraphics[width=  0.25\columnwidth]{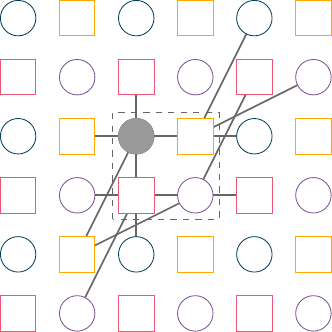}} 
\caption{The scenario when we use padding qubits. The left data qubit in the boxed basic unit is not part of the code, and act as a padding qubit. (a) Phase 1  (b) Phase 2, layer 3-4 (c) Phase 2 layer 5 (d) Phase 3 (e) The qubit connectivity assumed.}
\label{MQPad}
\end{figure}

In the example we present in Fig.~\ref{MQPad}, we see that the padding qubit is not part of the code, nevertheless we assumed that it shares the same connectivity as the other the basic units. In the first 4 layers, the ancilla do not interact with the padding qubit, since it is not part of any stabilizer. During the routing layer, a $\SWAP$ gate is manually inserted to ensure the net motion of the Z sublattice. Therefore, in Phase 3 of the syndrome extraction circuit, the Z-ancilla uses couplers connected to the padding qubit to perform the relevant two-qubit gates.

The second scenario is that such an padding qubit does not exist, in which case Louvre would fail to apply, but only locally. Some additional couplers will need to be added, connecting the static qubit to the qubits it will need to interact with afterwards. The added couplers are unlikely to point along any of the $v_{A_i,X}$, rendering the qubit connectivity on the hardware rather ugly. Nevertheless, the resulted number of couplers connecting the static qubit will not exceed the number of couplers required without applying Louvre. We imagine that this scenario will likely be the case when the hardware is specially designed to perform a certain code, in which case it should not be a problem to engineer ugly connectivity as such.

\begin{figure}[h]
\centering
\subfloat[]{\includegraphics[width=  0.25\columnwidth]{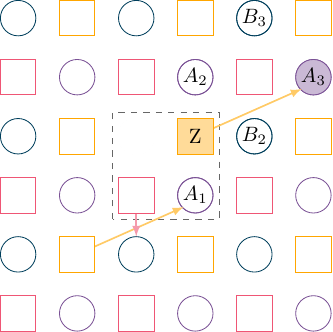}} \hspace{5mm} 
\subfloat[]{\includegraphics[width=  0.25\columnwidth]{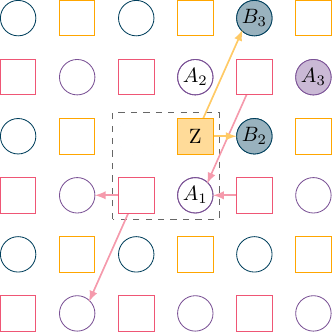}} \hspace{5mm} 
\subfloat[]{\includegraphics[width=  0.25\columnwidth]{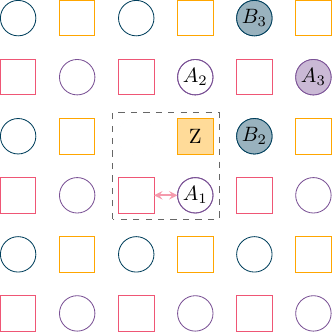}} \\
\subfloat[]{\includegraphics[width=  0.25\columnwidth]{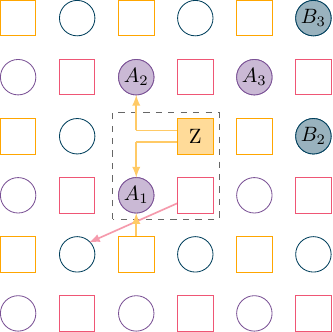}} \hspace{5mm} 
\subfloat[]{\includegraphics[width=  0.25\columnwidth]{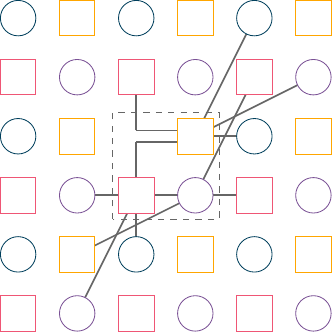}} 
\caption{The scenario when we install additional couplers. The left data qubit in the boxed basic unit is absent. (a) Phase 1  (b) Phase 2, layer 3-4 (c) Phase 2 layer 5 (d) Phase 3 (e) The qubit connectivity required.}
\label{MQBrut}
\end{figure}

In the example we present in Fig.~\ref{MQBrut}, we see that during the routing layer, The Z-ancilla in the boxed basic unit remains static. Therefore, in Phase 3 of the syndrome extraction circuit,  two additional couplers need to be installed to connect the Z-ancilla to the corresponding data qubits $A_1$ and $A_2$. 

\begin{figure}[ht]
\centering
\includegraphics[width=  0.6
\columnwidth]{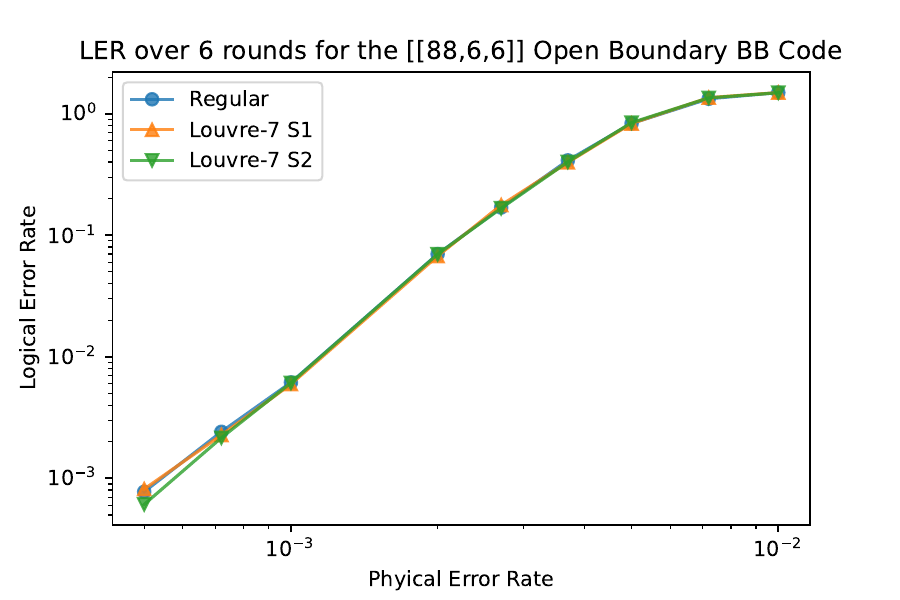} 
\caption{The logical error rate of the $\llbracket88,6,6\rrbracket$ BB on open boundary conditions with different syndrome extraction circuit. S1 is the first senario while S2 is the second senario discuss in this section.}
\label{OBBB_LER}
\end{figure}

We have simulated the logical error rate of a $\llbracket88,6,6\rrbracket$ BB code adapted to open boundary conditions \cite{liang_planar_2025}, and compared the logical error rate of the regular syndrome extraction circuit, Louvre-7 with the first scenario, and with the second scenario. As shown in Fig.~\ref{OBBB_LER}, all three circuits display similar logical error rates, indicating minor effect from the additional $\SWAP$ gates on the boundary in scenario 1.

\bibliographystyle{quantum}
\bibliography{qLDPC}
\end{document}